\documentclass{article}
\usepackage{caption}
\usepackage{subcaption}
\usepackage{authblk}
\usepackage{amsmath}
\usepackage{graphicx}
\usepackage{amssymb}
\usepackage{booktabs}
\usepackage{amsmath}
\usepackage{bm}
\usepackage{siunitx} % for \unit and \qty macros

\begin{document}

%% Version in square brackets will appear at top right of all pages except title page:
\title{Neural-Network Closures for Complex-Shaped Particles in the Force-Coupling Method}

\author[1]{Marco Laudato\footnote{Corresponding author; laudato@kth.se}}

\affil[1]{Department of Engineering Mechanics, FLOW Centre, KTH Royal Institute of Technology, Stockholm, SE-10044, Sweden.}

\maketitle

\begin{abstract}
A data-driven surrogate framework to accelerate particle-resolved modelling of quasi-dilute suspensions of rigid, non-spherical particles in Stokes flow is introduced.
A regularized-Stokeslet boundary element method (BEM) is implemented to compute hydrodynamic responses in canonical linear flows, focusing on the particle stresslet and angular velocity for spheroids, and additionally the chiral thrust for helicoidal particles. 
For spheroids, the BEM solver is validated against available analytical benchmarks (Faxén-type relations for the stresslet and Jeffery’s theory for rotation), and parameter choices for surface discretization and regularization are selected through systematic convergence studies.
For helicoidal particles, where no analytical solutions exist, accuracy is quantified via Richardson-style self-convergence, complemented by tests of linearity, frame objectivity, and chirality-dependent symmetries.
The resulting datasets are used to train a neural-operator surrogate that maps local flow descriptors and particle configuration to the corresponding stresslet, rotation, and thrust at negligible evaluation cost. Across independent test sets spanning random orientations and flow types, the surrogate achieves median relative errors below 1\% for the deviatoric stresslet (95th percentile below 3\%) and comparable accuracy for angular velocity and thrust.
The combination of validated BEM generation and fast inference provides a practical route to coupling complex particle shapes into mesoscale solvers such as the force-coupling method, enabling large-ensemble studies of microstructure and suspension rheology.

\end{abstract}
%  End of title page for Preprint option --------------------------------- %

\section{Introduction}
\label{sec:intro}
Understanding and predicting the rheology of particle-laden flows remains a central challenge in fluid mechanics, with applications ranging from industrial slurries and pharmaceutical formulations to microfluidic devices and biological suspensions~\cite{stickel2005fluid, mastropietro2013rheology,saintillan2018rheology}.
For dilute and quasi-dilute suspensions in the low Reynolds number regime, the force-coupling method (FCM) has emerged as the method of choice, enabling efficient simulation of thousands of particles by representing each particle as a regularized multipole distribution rather than resolving its detailed surface geometry~\cite{maxey2001localized}.
The resulting matrix-free mobility formulation scales almost linearly with particle count and has been successfully applied to suspensions of spheres, ellipsoids, and active swimmers in both periodic and confined domains~\cite{lomholt2003force,delmotte2015large}. This computational efficiency has made FCM a standard tool for systematic rheological studies, where understanding how suspension properties depend on particle shape, flow type, and volume fraction requires exploring high-dimensional parameter spaces with statistically converged ensemble averages~\cite{lomholt2002experimental, maxey2017simulation}.

However, FCM's efficiency comes at a cost: it requires an analytical or semi-analytical closure relating the local velocity gradient around each particle to its hydrodynamic stresslet, i.e. the dipole force distribution that governs suspension stress and particle motion.
For spheres and axisymmetric ellipsoids, such closures follow directly from classical microhydrodynamics.
Faxén relations provide exact expressions for the force, torque, and stresslet of rigid spheres and spheroids in arbitrary linear flows~\cite{rallison1978note}, while Jeffery and Bretherton theories describe the orientation dynamics and stress contributions in shear~\cite{bretherton1962motion}.
Comprehensive tables of resistance and mobility tensors for such simple shapes are available in standard references~\cite{kim2013microhydrodynamics}, and these have been successfully embedded into FCM with appropriate rescaling of the regularized envelopes~\cite{liu2009force}.
But for particles with genuinely complex three-dimensional geometry (e.g., helical chiral bodies, toroidal shapes, rough or faceted surfaces, or multi-scale aggregates), deriving the required tensorial closures would demand bespoke boundary element or finite element calculations for each new shape, followed by curve-fitting procedures to extract functional forms suitable for real-time evaluation during time-stepping.
In practice, this reliance on shape-specific analytical closures effectively restricts current FCM implementations to spheres and spheroids, preventing systematic investigations of how three-dimensional particle geometry controls suspension rheology.

The alternative is to use fully resolved methods that discretize the particle surface explicitly.
Interface-resolved approaches such as the immersed boundary method (IBM), lattice Boltzmann method (LBM), and high-fidelity boundary element or finite element formulations can treat arbitrary particle shapes with high accuracy, capturing lubrication interactions, surface roughness, and detailed flow structures near complex boundaries~\cite{ladd2001lattice, saadat2018immersed}.
Yet these methods come with substantial computational cost: IBM and related Lagrangian–Eulerian coupling schemes require fine grid resolution near particle surfaces, with costs scaling at least linearly in grid size and often with large prefactors due to the need for resolving thin lubrication gaps~\cite{yu2010lattice}.
LBM-based particle simulations similarly require uniform fine lattices and typically restrict three-dimensional studies to $\mathcal{O}(10^2-10^3)$ particles~\cite{aidun2010lattice,bagge2021highly}.
Boundary integral and boundary element methods (BEM) for Stokes flow offer excellent single-particle accuracy but require assembling and solving dense linear systems whose size grows with surface discretization, making many-particle calculations increasingly prohibitive~\cite{yeo2010simulation,zhao2023role}.
As a result, while fully resolved methods are indispensable for generating benchmark data and for detailed studies of small suspensions, their computational expense limits systematic rheological studies (such as mapping intrinsic viscosity across shape parameters, exploring transient orientation dynamics, or computing statistically converged suspension stress in different flow types) to configurations with at most a few hundred particles.
Even for quasi-dilute suspensions (volume fractions $\phi\approx0.10-0.15$) the cost of systematic exploration over shape, flow, and concentration studies with well-converged ensemble statistics (with particles number of $\mathcal{O}(10^4)$) makes this approach unpractical.

This computational bottleneck has left fundamental questions about complex-shaped particle suspensions largely unanswered.
Even for simple canonical flows, we lack systematic rheological state diagrams for suspensions of helicoidal or toroidal particles, i.e. geometries that cannot be approximated by spheroids and for which no analytical Faxén relations exist, despite their relevance for chiral colloids, helical microswimmers, enantioselective separation, and functional materials with engineered anisotropy.
Similarly, the effect of surface roughness, faceting, or chirality on suspension viscosity, normal-stress differences, and microstructural development remains under-explored, often restricted to dilute limits or highly idealized models.

At the same time, recent efforts to accelerate fluid–particle simulations using machine learning demonstrate both the community's recognition of this computational challenge and the opportunity to develop data-driven closures.
However, these ML approaches have so far focused on learning many-body mobilities from suspension-scale data~\cite{ma2025h,howard2023machine}, inferring continuum constitutive models from macroscopic measurements~\cite{wang2021machine,reyes2021learning}, or replacing entire solvers with neural operators trained on fully resolved simulations~\cite{laudato2025neuralmdpi,laudato2024high,laudato2025neural}.
None provide a shape-resolving single-particle closure that can be embedded directly into FCM to enable systematic rheological studies of arbitrary three-dimensional geometries at the quasi-dilute concentrations where microstructural effects emerge but single-particle hydrodynamics still dominate the stress response.

We address this gap by introducing a neural-network surrogate that acts as a general, shape-resolving closure for FCM, applicable to \textit{arbitrarily shaped} rigid particles in the Stokes flow regime.
The key idea is to decouple the expensive geometric calculation from the many-particle simulation: we compute accurate single-particle hydrodynamic responses offline using (for example) boundary element simulations for the particle shape of interest, sampling a comprehensive dataset spanning orientation angles and canonical flow types (shear, uniaxial extension, planar extension, biaxial extension).
We then train a neural surrogate model to approximate the nonlinear closure mapping the local linearized flow and particle orientation to the particle's hydrodynamic response.
Concretely, the surrogate learns

\begin{equation}
    \label{eq:surrogate}
    \left(\mathbf{E},\mathbf{\Omega},\mathbf{p}\right)\longrightarrow\left(\mathbf{S}, \mathbf{\omega},\mathbf{F}_p\right)
\end{equation}

\noindent
where $\mathbf{E}$ and $\mathbf{\Omega}$ are the strain-rate and vorticity tensors of the local linearized flow ($\nabla\mathbf{u}_{loc}=\mathbf{E}+\mathbf{\Omega}$), $\mathbf{p}$ encodes the particle orientation, $\mathbf{S}$ is the stresslet tensor, $\mathbf{\omega}$ is the rigid-body angular velocity, and $\mathbf{F}_p$ is a propulsive thrust (relevant if dealing with helical particles).

To exemplify the potential of the method, we will implement  neural surrogate models for spheroids and helicoidal particles, whereas the corresponding training dataset is built using boundary element method to compute the single-particle hydrodynamic responses. The choice of helicoidal chiral particles as a test case is deliberate.
Unlike spheroids, for which analytical solutions provide validation benchmarks, helical particles possess no analytical Faxén relations and exhibit genuinely three-dimensional, chirality-dependent hydrodynamics: their stresslet response is even under parity inversion while their propulsive thrust is odd, coupling to the local flow through terms that depend explicitly on the handedness~\cite{ishimoto2020helicoidal}.
This makes them an ideal challenge problem for demonstrating that the surrogate framework can learn complex, symmetry-constrained physics from data alone.
Moreover, chiral suspensions have received limited systematic study despite applications in enantioselective separation~\cite{meinhardt2012separation}, microfluidic sorting of helical bacteria~\cite{marcos2009separation}, and self-assembly of chiral colloids~\cite{parton2022chiral}, precisely because existing methods cannot efficiently simulate thousands of complex three-dimensional chiral particles across parameter space.

Once trained, this surrogate provides a drop-in closure for FCM: at each timestep and for each particle, the FCM solver supplies the local flow gradient and orientation to the neural network and receives the stresslet, rotation rate, and thrust needed to advance the particle dynamics and update the suspension stress.
Critically, the computational cost of evaluating the trained surrogate (a single forward pass through a network with $\sim30,000 - 80,000$  parameters) is negligible compared to both the FCM fluid solve and the original BEM calculation, yielding major speedups over direct boundary element methods while maintaining sub-1\% median relative error in stresslet and angular velocity predictions.
This transforms the cost structure: the expensive boundary element calculations are confined to a one-time offline training stage (reusable across all subsequent simulations), while online many-particle FCM simulations retain their characteristic near-linear scaling with particle count, now extended to arbitrary three-dimensional shapes.

Our framework enables systematic exploration of suspension rheology for complex-shaped particles: questions such as whether handedness affects bulk viscosity, how chiral particle orientation correlates with flow type, and how thrust coupling modifies suspension stress, (which we defer to future work) are now computationally tractable.

In this work, we develop and systematically validate this neural-network closure framework for rigid, neutrally buoyant, non-Brownian particles in quasi-dilute suspensions ($\phi\approx0.10-0.15$) under Stokes flow conditions.
We focus on this regime because single-particle hydrodynamics dominate the stress response and can be computed accurately with BEM, providing clean training data and enabling rigorous validation against analytical benchmarks where available.
Extensions to higher concentrations (lubrication corrections, contact models), finite Reynolds number , active matter (self-propulsion), and deformable particles (fluid-structure interaction) are natural next steps, but the dilute baseline established here is essential for validating the surrogate methodology before introducing additional physical complexity.

To generate the training dataset, we implement a regularized Stokeslet boundary element solver for axisymmetric spheroids and three-dimensional helicoidal chiral particles, validate the BEM implementation against analytical solutions and using Richardson extrapolation, and finally generate comprehensive datasets spanning multiple orientations and flow types.
We then train fully connected neural-network surrogates with physics-informed input features (strain invariants, Jeffery-orbit terms) for both spheroidal and helicoidal geometries.
We conduct ablation studies comparing network architectures (moderate vs. large), feature engineering strategies (raw inputs vs. physics-informed features), and data augmentation techniques (chirality mirroring for helical particles), quantifying surrogate accuracy on held-out test data and estimating error propagation when the model is embedded in an FCM solver.
Across all tested configurations, the best-performing models achieve median relative errors $\leq1\%$ for deviatoric stresslet components and $\approx 1.3\%$ for angular velocity, indicating that the proposed surrogate can serve as a highly accurate and computationally efficient closure for systematic rheological studies.

The article is organized as follows.
Section~\ref{sec:single_particle} describes the boundary element method implementation for both spheroids and helicoidal particles, including regularized Stokeslet formulations, surface discretization strategies, validation against analytical solutions, and Richardson extrapolation for error estimation.
Section~\ref{sec:Surrogate} presents the neural network architecture, physics-informed feature engineering, training methodology with optional physics-based loss functions, and chirality data augmentation for helical particles.
Section~\ref{sec:accuracy} quantifies surrogate accuracy through ablation studies, and estimates how surrogate errors propagate to macroscopic suspension stress when embedded in FCM.
Section~\ref{sec:conclusions} concludes with the limitations of the current framework (quasi-dilute regime, Stokes flow), and pathways for future directions.

\section{Particle-Scale Hydrodynamics}
\label{sec:single_particle}
\subsection{Single-particle Stokes problem in a linear flow}
\label{ssec:stokes_problem}
In the quasi-dilute regime, the hydrodynamic interaction between suspended particles can be captured by the response of a single rigid particle to a locally linear flow. We consider an isolated particle with surface $\partial\mathcal B$ immersed in an incompressible Newtonian fluid of viscosity $\mu$ at zero Reynolds number. The local macroscopic flow field $\bar{\boldsymbol u}(\boldsymbol x)$ is linearized around a reference point $\boldsymbol x_c$ (typically the particle centroid) as
\begin{equation}
    \boldsymbol u_\infty(\boldsymbol x) 
    = \bar{\boldsymbol u}(\boldsymbol x_c) 
    + \boldsymbol A\left(\boldsymbol x - \boldsymbol x_c\right),
    \qquad 
    \boldsymbol A = \nabla \bar{\boldsymbol u}(\boldsymbol x_c),
\end{equation}
where $\boldsymbol u_\infty(x)$ denotes the imposed (undisturbed) background velocity field
in the neighbourhood of the particle and $\boldsymbol A$ is the imposed velocity gradient. It is convenient~\cite{leal2007advanced} to decompose $\boldsymbol A$ into its symmetric and antisymmetric parts,
\begin{equation}
    \boldsymbol E = \tfrac12\left(\boldsymbol A + \boldsymbol A^\mathsf T\right), 
    \qquad
    \boldsymbol W = \tfrac12\left(\boldsymbol A - \boldsymbol A^\mathsf T\right),
\end{equation}
where $\boldsymbol E$ is the rate-of-strain tensor and $\boldsymbol W$ represents the rigid-body rotation of the background flow.

The total velocity field $\boldsymbol u$ and pressure $p$ satisfy the steady Stokes equations in the fluid domain exterior to the particle,
\begin{equation}
    -\nabla p + \mu \nabla^2 \boldsymbol u = \boldsymbol 0, 
    \qquad 
    \nabla \cdot \boldsymbol u = 0,
\end{equation}
subject to the no-slip condition on the particle surface and the matching to the imposed flow at infinity. The particle undergoes an unknown rigid-body translation $\boldsymbol U$ and rotation $\boldsymbol\Omega$, so that the no-slip boundary condition can be written as~\cite{kim2013microhydrodynamics}
\begin{equation}
    \boldsymbol u(\boldsymbol x) 
    = \boldsymbol u_\infty(\boldsymbol x) 
    + \boldsymbol U 
    + \boldsymbol\Omega \times \left(\boldsymbol x - \boldsymbol x_c\right),
    \qquad \boldsymbol x \in \partial\mathcal B,
\end{equation}
while the disturbance decays far from the particle,
\begin{equation}
    \boldsymbol u(\boldsymbol x) - \boldsymbol u_\infty(\boldsymbol x) \to \boldsymbol 0 
    \quad \text{as} \quad \lvert\boldsymbol x\rvert \to \infty.
\end{equation}

We assume neutrally buoyant particles and no external body forces or torques. The hydrodynamic traction on the surface is 
$\boldsymbol f = \boldsymbol\sigma \boldsymbol n$, where $\boldsymbol n$ is the outward normal and the Cauchy stress tensor is
\begin{equation}
    \boldsymbol\sigma(\boldsymbol u, p) 
    = -p\,\boldsymbol I + 2\mu\,\boldsymbol E(\boldsymbol u), 
    \qquad
    \boldsymbol E(\boldsymbol u) 
    = \tfrac12\left(\nabla \boldsymbol u + \nabla \boldsymbol u^\mathsf T\right).
\end{equation}
Force- and torque-free motion then implies the global constraints
\begin{equation}
    \boldsymbol F 
    = \int_{\partial\mathcal B} \boldsymbol f \,\mathrm dS 
    = \boldsymbol 0, 
    \qquad
    \boldsymbol T 
    = \int_{\partial\mathcal B} 
    \left(\boldsymbol x - \boldsymbol x_c\right) \times \boldsymbol f \,\mathrm dS 
    = \boldsymbol 0,
\end{equation}
which determine $\boldsymbol U$ and $\boldsymbol\Omega$ for a given imposed velocity gradient $\boldsymbol A$.

The quantity that links this single-particle problem to the macroscopic rheology of a suspension is the \emph{stresslet}, defined as the symmetric first moment of the surface traction,
\begin{equation}
    \boldsymbol S 
    = \frac12 \int_{\partial\mathcal B} 
    \big[
    \left(\boldsymbol x - \boldsymbol x_c\right) \otimes \boldsymbol f
    + \boldsymbol f \otimes \left(\boldsymbol x - \boldsymbol x_c\right)
    \big] \,\mathrm dS.
\end{equation}
In incompressible suspensions it is customary to consider the deviatoric part 
$\boldsymbol S^{\mathrm{dev}}$, obtained by subtracting one third of the trace times the identity. For a dilute suspension of identical particles with number density $n$, the bulk extra stress is
\begin{equation}
    \boldsymbol\Sigma^{\mathrm{p}} 
    = n\,\langle \boldsymbol S^{\mathrm{dev}} \rangle,
\end{equation}
where $\langle \cdot \rangle$ denotes an ensemble average over particle positions and orientations. The total macroscopic stress is then written as~\cite{batchelor1970stress}
\begin{equation}
    \boldsymbol\Sigma 
    = -\bar{p}\,\boldsymbol I 
    + 2\mu\,\boldsymbol E_\infty 
    + \boldsymbol\Sigma^{\mathrm{p}},
\end{equation}
with $\boldsymbol E_\infty$ the imposed macroscopic rate-of-strain. This relation is valid for arbitrary rigid particle shapes in the dilute limit; the shape only enters through the dependence of $\boldsymbol S$ on $\boldsymbol E_\infty$ and on the particle orientation.

In a simple shear flow with shear rate $\dot\gamma$ (so that $E_{\infty,xy} = \dot\gamma/2$), a scalar effective viscosity $\mu_\mathrm{eff}$ can be defined from the macroscopic shear stress,
\begin{equation}
    \mu_\mathrm{eff} 
    = \frac{\Sigma_{xy}}{\dot\gamma}
    = \mu + \frac{n\,\langle S^{\mathrm{dev}}_{xy} \rangle}{\dot\gamma}.
\end{equation}
For rigid spheres, inserting the analytical stresslet into this expression recovers Einstein's effective viscosity 
$\mu_\mathrm{eff} = \mu\left(1 + 2.5\phi\right)$, where $\phi$ is the volume fraction~\cite{einstein1911berichtigung}. For non-spherical or chiral particles, the same definition generally leads to anisotropic effective viscosities and non-zero normal-stress differences, all encoded in the tensorial structure of $\boldsymbol S$.

The stresslet $\boldsymbol S$ and the associated rigid-body motion $(\boldsymbol U, \boldsymbol\Omega)$ provide the microscale closure that is required by the force-coupling method at the suspension scale. We therefore focus in the following on an accurate and efficient evaluation of $\boldsymbol S$ for particles of arbitrary shape in a linear flow, using a boundary element formulation of the Stokes problem introduced above.

\subsection{Boundary element formulation}
\label{ssec:BEM}
To solve the single-particle Stokes problem introduced above, we use a boundary element formulation based on the free-space Green's function of the Stokes equations. The bulk Stokes equations are satisfied identically by construction, and the numerical problem reduces to enforcing the no-slip condition on the particle surface together with the global force- and torque-free constraints.

\subsubsection{Regularized Stokeslet representation}

We decompose the total velocity as $\boldsymbol u = \boldsymbol u_\infty + \boldsymbol u'$, where $\boldsymbol u_\infty(\boldsymbol x) = \boldsymbol A(\boldsymbol x-\boldsymbol x_c)$ is the imposed linear flow and $\boldsymbol u'$ is the disturbance induced by the particle. The disturbance is represented as a single-layer potential of regularized Stokeslets~\cite{pozrikidis1992boundary}
\begin{equation}
    \boldsymbol u'(\boldsymbol x)
    = \int_{\partial\mathcal B} \boldsymbol G_\varepsilon(\boldsymbol x-\boldsymbol y)\,
    \boldsymbol f(\boldsymbol y)\,\mathrm dS(\boldsymbol y),
\end{equation}
where $\boldsymbol f = \boldsymbol\sigma\boldsymbol n$ is the surface traction and $\boldsymbol G_\varepsilon$ is a regularized version of the free-space Stokeslet.

The (singular) \emph{Stokeslet} is the Green's function of the Stokes equations, i.e.\ the velocity field generated by a unit point force applied at the origin in an unbounded viscous fluid. In components, the free-space Stokeslet $\boldsymbol G$ reads
\begin{equation}
    G_{ij}(\boldsymbol r)
    = \frac{1}{8\pi\mu}
    \left(
        \frac{\delta_{ij}}{r}
        + \frac{r_i r_j}{r^3}
    \right),
    \qquad
    \boldsymbol r = \boldsymbol x - \boldsymbol y,
    \quad r = \lvert\boldsymbol r\rvert,
\end{equation}
and provides the fundamental singular solution to the Stokes equations. In numerical computations we instead employ a regularized Stokeslet $\boldsymbol G_\varepsilon$ of Cortez type~\cite{cortez2018regularized}, 
\begin{equation}
    G_{\varepsilon,ij}(\boldsymbol r)
    = \frac{1}{8\pi\mu}
    \left[
        \frac{r^2 + 2\varepsilon_{reg}^2}{(r^2 + \varepsilon_{reg}^2)^{3/2}}\,\delta_{ij}
        + \frac{r_i r_j}{(r^2 + \varepsilon_{reg}^2)^{3/2}}
    \right],
    \label{eq:reg_stokeslet}
\end{equation}
which recovers the classical Stokeslet as $\varepsilon_{reg}\to 0$. The regularization length $\varepsilon_{reg}$ is chosen proportional to the local surface spacing, so that the kernel remains bounded as $\boldsymbol x\to\boldsymbol y$ while retaining the correct far-field behaviour.
For a surface discretized with $N$ nodes and of total area $A_p$, we set $\varepsilon_{reg}=\varepsilon\sqrt{A_p/N}$, where $\varepsilon$ is an $\mathcal{O}(1)$ prefactor (typically $\varepsilon\approx0.4-0.5$). This choice ensures that $\varepsilon_{reg}$ remains small compared to the particle size while being large enough to regularize the kernel at coinciding collocation points.

On the particle surface we impose no slip with rigid-body motion,
\begin{equation}
    \boldsymbol u(\boldsymbol x)
    = \boldsymbol U + \boldsymbol\Omega\times(\boldsymbol x-\boldsymbol x_c),
    \qquad \boldsymbol x\in\partial\mathcal B.
\end{equation}
Combining this with the integral representation yields, at any point $\boldsymbol x\in\partial\mathcal B$,
\begin{equation}
    \boldsymbol u_\infty(\boldsymbol x)
    + \int_{\partial\mathcal B} \boldsymbol G_\varepsilon(\boldsymbol x-\boldsymbol y)\,
    \boldsymbol f(\boldsymbol y)\,\mathrm dS(\boldsymbol y)
    = \boldsymbol U + \boldsymbol\Omega\times(\boldsymbol x-\boldsymbol x_c).
\end{equation}
In practice we recenter the discretized surface so that the quadrature centroid coincides with the origin, $\boldsymbol x_c = \boldsymbol 0$, which simplifies the rigid-body contribution to $\boldsymbol U + \boldsymbol\Omega\times\boldsymbol x$.

\subsubsection{Surface discretization and numerical quadrature}

The particle surface $\partial\mathcal B$ is discretized by a set of $N$ nodes $\{\boldsymbol x_j\}_{j=1}^N$ and associated quadrature weights $\{w_j\}_{j=1}^N$. For any scalar or vector field $g$ defined on $\partial\mathcal B$, surface integrals are approximated as
\begin{equation}
    \int_{\partial\mathcal B} g(\boldsymbol y)\,\mathrm dS(\boldsymbol y)
    \approx \sum_{j=1}^N w_j\,g(\boldsymbol x_j).
\end{equation}
The weights $w_j$ can be interpreted as the effective area associated with node $j$, chosen so that the quadrature is accurate for smooth fields $g$.

For spheroids, we first generate an almost-uniform Fibonacci distribution on the unit sphere and then map the nodes affinely to the spheroidal surface; the Jacobian of this mapping provides the corresponding area weights. For helical particles, the surface is generated by sweeping a circular cross-section of fixed wire radius along a helical centreline, with the local Frenet frame used to orient the cross-section; the weights follow from the local tube geometry. In both cases the discretized surface is translated so that the weighted centroid is at the origin and rotated to align a prescribed particle axis with the desired direction in the laboratory frame.

With this quadrature rule, the boundary integral in the Stokeslet representation is approximated as
\begin{equation}
    \int_{\partial\mathcal B} \boldsymbol G_\varepsilon(\boldsymbol x_i-\boldsymbol y)\,
    \boldsymbol f(\boldsymbol y)\,\mathrm dS(\boldsymbol y)
    \approx
    \sum_{j=1}^N w_j\,\boldsymbol G_\varepsilon(\boldsymbol x_i-\boldsymbol x_j)\,\boldsymbol f_j,
\end{equation}
where $\boldsymbol f_j$ denotes the traction at node $j$. Evaluated at the collocation points $\boldsymbol x_i$ on the surface (which we take to be the same as the quadrature nodes), the no-slip boundary condition becomes
\begin{equation}
    \sum_{j=1}^N w_j\,\boldsymbol G_\varepsilon(\boldsymbol x_i-\boldsymbol x_j)\,\boldsymbol f_j
    - \boldsymbol U
    - \boldsymbol\Omega\times\boldsymbol x_i
    = -\boldsymbol u_\infty(\boldsymbol x_i),
    \qquad i=1,\dots,N.
\end{equation}

\subsubsection{Linear system and global constraints}

We collect all nodal tractions in a single vector
$\boldsymbol f = (\boldsymbol f_1,\dots,\boldsymbol f_N)\in\mathbb R^{3N}$ and form the dense matrix $\boldsymbol G\in\mathbb R^{3N\times 3N}$ with $3\times3$ blocks
\begin{equation}
    \boldsymbol G_{ij} = w_j\,\boldsymbol G_\varepsilon(\boldsymbol x_i-\boldsymbol x_j),
    \qquad i,j=1,\dots,N.
\end{equation}
The discretized no-slip conditions can then be written compactly as
\begin{equation}
    \boldsymbol G\,\boldsymbol f
    - \boldsymbol U_{\mathrm{blk}}
    - \boldsymbol C\,\boldsymbol\Omega
    = -\boldsymbol u_\infty,
\end{equation}
where $\boldsymbol u_\infty\in\mathbb R^{3N}$ stacks the imposed velocity at all nodes, $\boldsymbol U_{\mathrm{blk}}$ is the block vector that repeats $\boldsymbol U$ at each node, and $\boldsymbol C$ is the block matrix built from the cross-product matrices $[\boldsymbol x_i]_\times$, so that $\boldsymbol C\,\boldsymbol\Omega$ represents $\boldsymbol\Omega\times\boldsymbol x_i$ at each node.

The global force- and torque-free conditions are discretized consistently with the same quadrature rule,
\begin{equation}
    \boldsymbol F \approx \sum_{j=1}^N w_j\,\boldsymbol f_j = \boldsymbol 0,
    \qquad
    \boldsymbol T \approx \sum_{j=1}^N w_j\,\boldsymbol x_j\times\boldsymbol f_j = \boldsymbol 0,
\end{equation}
and are expressed as
\begin{equation}
    \boldsymbol F_{\mathrm{rows}}\,\boldsymbol f = \boldsymbol 0, 
    \qquad
    \boldsymbol T_{\mathrm{rows}}\,\boldsymbol f = \boldsymbol 0,
\end{equation}
where $\boldsymbol F_{\mathrm{rows}}, \boldsymbol T_{\mathrm{rows}}\in\mathbb R^{3\times 3N}$ are block matrices with
$\boldsymbol F_{\mathrm{rows}} = [w_1 I_3 \;\; \dots \;\; w_N I_3]$ and
$\boldsymbol T_{\mathrm{rows}} = [w_1[\boldsymbol x_1]_\times \;\; \dots \;\; w_N[\boldsymbol x_N]_\times]$.
Here $[\boldsymbol x_j]_\times$ denotes the $3\times 3$ skew-symmetric matrix associated with the cross product by $\boldsymbol x_j = (x_{j,1},x_{j,2},x_{j,3})^\mathsf T$,
\begin{equation}
    [\boldsymbol x_j]_\times
    =
    \begin{bmatrix}
        0        & -x_{j,3} &  x_{j,2} \\
        x_{j,3}  &  0       & -x_{j,1} \\
       -x_{j,2}  &  x_{j,1} &  0
    \end{bmatrix},
\end{equation}
so that $[\boldsymbol x_j]_\times \boldsymbol f_j = \boldsymbol x_j \times \boldsymbol f_j$.

Combining the no-slip conditions with these constraints yields the linear system for the unknowns $(\boldsymbol f,\boldsymbol U,\boldsymbol\Omega)$,
\begin{equation}
    \begin{bmatrix}
        \boldsymbol G & -\boldsymbol I & -\boldsymbol C \\
        \boldsymbol F_{\mathrm{rows}} & \boldsymbol 0 & \boldsymbol 0 \\
        \boldsymbol T_{\mathrm{rows}} & \boldsymbol 0 & \boldsymbol 0
    \end{bmatrix}
    \begin{bmatrix}
        \boldsymbol f \\[2pt] \boldsymbol U \\[2pt] \boldsymbol\Omega
    \end{bmatrix}
    =
    \begin{bmatrix}
        -\boldsymbol u_\infty \\[2pt] \boldsymbol 0 \\[2pt] \boldsymbol 0
    \end{bmatrix}.
\end{equation}
This corresponds to a passive particle that is free to translate and rotate in response to the imposed linear flow. 

From a computational viewpoint, the boundary element formulation leads to a dense linear system of size $3N\times 3N$, where $N$ is the number of surface nodes.
Forming the dense regularized Stokeslet matrix $\boldsymbol{G}$ by evaluating all pairwise kernel interactions scales as $\mathcal{O}(N^2)$ operations and requires $\mathcal{O}(N^2)$ storage, while solving the resulting $N \times N$ system with a direct dense method such as Gaussian elimination or LU factorization requires $\mathcal{O}(N^3)$ floating-point operations and $\mathcal{O}(N^2)$ memory~\cite{Wang2018ATA}.
While this is acceptable for computing a single high-fidelity solution, it rapidly becomes prohibitive when one needs stresslets and rigid-body motion for many combinations of particle shapes, orientations and imposed linear flows, or when such evaluations would have to be performed repeatedly at every time step and for every particle in a suspension simulation.
This unfavorable scaling motivates the use of a surrogate model that emulates the BEM response at a drastically reduced online cost, while still retaining the accuracy of the underlying boundary element solutions used for training.

We implemented the proposed boundary element solver in Python, using a fully modular workflow for geometry generation, quadrature, assembly, and linear solves. The complete implementation, together with scripts to reproduce the validation studies and figures reported in this work, is openly available at~\cite{laudato2025_neuralfcm_github}.

\subsection{Linearity and scaling of the single-particle response}
\label{ssec:linearity}
The boundary element formulation described above inherits the linear structure of the Stokes equations. For a given particle shape and orientation, the discrete problem is linear in the unknown traction, translation, and rotation, and the right-hand side depends linearly on the imposed velocity gradient. Together with the dimensional structure of Stokes flow, this leads to simple and useful scaling relations for the stresslet and rigid-body motion.

\subsubsection{Linearity in velocity gradient and viscosity}

At the discrete level, the BEM system has the schematic form
\begin{equation}
    \mathcal A(\mu, \text{geometry})
    \begin{bmatrix}
        \boldsymbol f \\ \boldsymbol U \\ \boldsymbol\Omega
    \end{bmatrix}
    =
    \mathcal B(\text{geometry}) : \boldsymbol A,
\end{equation}
where $\boldsymbol A$ is the imposed velocity gradient. The matrices $\mathcal A$ and $\mathcal B$ are independent of $\boldsymbol A$, and $\mathcal A$ depends on the viscosity only through the regularized Stokeslet kernel. As a consequence, if the velocity gradient is scaled as $\boldsymbol A \rightarrow \alpha \boldsymbol A$, the solution scales linearly,
\begin{equation}
    \boldsymbol f \rightarrow \alpha \boldsymbol f, \qquad
    \boldsymbol U \rightarrow \alpha \boldsymbol U, \qquad
    \boldsymbol\Omega \rightarrow \alpha \boldsymbol\Omega.
\end{equation}
If the viscosity is scaled as $\mu \rightarrow \beta \mu$ while keeping the imposed velocity field fixed, the kinematics $\boldsymbol U$ and $\boldsymbol\Omega$ remain unchanged, whereas the tractions scale as $\boldsymbol f \rightarrow \beta \boldsymbol f$.

The stresslet, being a linear functional of the traction, therefore obeys
\begin{equation}
    \boldsymbol S \propto \mu\,\boldsymbol A,
\end{equation}
for fixed particle geometry and orientation. Equivalently, in terms of the rate-of-strain tensor $\boldsymbol E = (\boldsymbol A + \boldsymbol A^\mathsf T)/2$,
\begin{equation}
    \boldsymbol S \propto \mu\,\boldsymbol E.
\end{equation}
In a simple shear flow with shear rate $\dot\gamma$ (so that $E_{xy} = \dot\gamma/2$), this reduces to the familiar scaling $\boldsymbol S \propto \mu \dot\gamma$, consistent with the classical analytical result for a rigid sphere~\cite{batchelor1970stress,kim2013microhydrodynamics}.

\subsubsection{Geometric scaling of stresslet, rotation and thrust}

The dependence on particle size follows from dimensional analysis. Let $\ell$ be a characteristic particle length (for example, the radius of a sphere) and let $\gamma$ denote a characteristic strain rate associated with the imposed linear flow (e.g.\ $\gamma = \|\boldsymbol E\|$). Introducing dimensionless variables based on $\ell$ and $\gamma$ removes all explicit reference to $\mu$, $\gamma$ and $\ell$ from the Stokes equations. The resulting dimensionless single-particle problem is fully determined by the shape, orientation, and type of linear flow.

For geometrically similar particles (i.e., related by scale transformations), this implies that:

\begin{itemize}
    \item the dimensionless stresslet depends only on shape, orientation and flow type, and the dimensional stresslet scales as
    \begin{equation}
        \boldsymbol S \sim \mu\,\gamma\,\ell^3,
    \end{equation}
    or, equivalently,
    \begin{equation}
        \boldsymbol S \propto \mu\,\gamma\,V,
    \end{equation}
    with $V \propto \ell^3$ the particle volume;

    \item the translational velocity scales as
    \begin{equation}
        \boldsymbol U \sim \gamma\,\ell,
    \end{equation}
    including any net drift along the particle axis for chiral particles in shear;

    \item the angular velocity scales as
    \begin{equation}
        \boldsymbol\Omega \sim \gamma,
    \end{equation}
    independent of $\mu$ and $\ell$ for fixed shape and orientation.
\end{itemize}

The rotation matrix describing the particle orientation is dimensionless and evolves according to $\dot R = \boldsymbol\Omega \times R$, so its dynamics depend only on the dimensionless shear, not on the absolute viscosity or particle size.

\subsubsection{Tensorial representation and connection to the surrogate}

These properties can be summarized by introducing a dimensionless fourth-order tensor $\mathsf M$ that encodes the single-particle response,
\begin{equation}
    \boldsymbol S(\boldsymbol E; \mu, \ell, \text{shape}, \text{orientation})
    = \mu\,\ell^3\,
    \mathsf M(\text{shape},\text{orientation}) : \boldsymbol E,
\end{equation}
where $\boldsymbol E$ is the imposed rate-of-strain tensor. For a sphere, $\mathsf M$ reduces to a scalar multiple of the identity acting on $\boldsymbol E$; for non-spherical or chiral particles, $\mathsf M$ encodes shape-induced anisotropy and chiral couplings. Analogous, lower-order tensors relate $\boldsymbol E$ to the translational and angular velocities, which in physical units scale as $\boldsymbol U \sim \gamma \ell$ and $\boldsymbol\Omega \sim \gamma$.

In the dilute limit of the force-coupling method, the tensor $\mathsf M$ plays the role of a microscopic mobility object: once $\mathsf M$ is known for a given particle shape and orientation, the contribution of that particle to the macroscopic stress follows directly, in combination with an appropriate orientation distribution~\cite{batchelor1970stress,kim2013microhydrodynamics}.
In this sense, our neural surrogate can be interpreted as a data-driven approximation of $\mathsf M$ (and of its counterparts for translation and rotation), conditioned on particle orientation and geometry.
We train the surrogate on dimensionless BEM data, where the trivial scalings with $\mu$, particle size and strain rate have been factored out, so that the network learns only the non-trivial dependence on shape, orientation and flow type.
Physical stresslets, velocities and rotations are then recovered in the force-coupling simulations by simple rescaling according to the relations above.

\subsection{Validation of the BEM models}
\label{ssec:validation_BEM}
\begin{figure}[h!]
    \centering
    \includegraphics[width=0.49\linewidth]{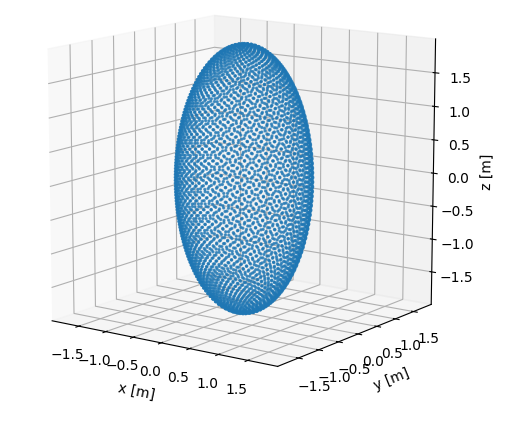}
    \includegraphics[width=0.49\linewidth]{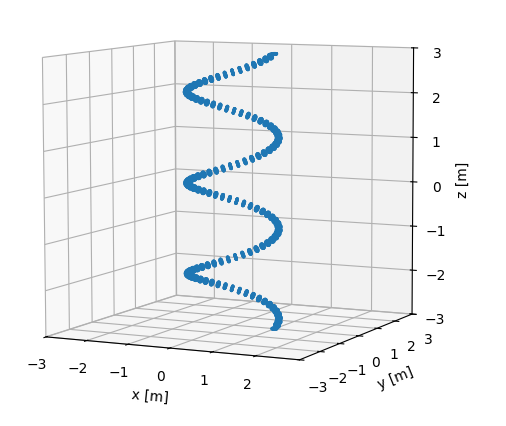}
    \caption{Discretized particle geometries used to train the surrogate. Left: spheroid surface mesh (boundary-element nodes) for the axisymmetric ellipsoid. Right: helicoidal particle surface discretization (boundary-element nodes), showing the chiral geometry used in the BEM simulations.}
    \label{fig:geometries}
\end{figure}

In this study we consider two representative particle geometries: axisymmetric spheroids and three-dimensional helicoidal (chiral) particles (Fig.~\ref{fig:geometries}).
This subsection is to validate the BEM solver accuracy for each geometry before generating the datasets used to train the surrogate models.

\subsubsection*{Validation for spheroids}
We validate the boundary element solver for an ellipsoid of revolution with semi-axes \((a,a,c)\) and aspect ratio \(r=c/a=2\).
The validation set consists of 32 cases obtained by combining 8 particle orientations with four linear flow types (simple shear, uniaxial, planar and biaxial extension).
For each case we compare the BEM deviatoric stresslet and Jeffery angular velocity with analytical spheroid solutions.
Analytical reference values are computed following the classical resistance formulation for ellipsoids in Stokes flow \cite{kim2013microhydrodynamics}: given \(r\), we evaluate the shape-dependent scalar resistance coefficients and assemble the fourth-order resistance tensor, whose contraction with the imposed rate-of-strain tensor \(\mathbf{E}\) yields the deviatoric stresslet \(\mathbf{S}^{\mathrm{an}}\)
The analytical angular velocity is obtained from Jeffery's equation \cite{jeffery1922motion},
\[
\boldsymbol{\Omega}_{\mathrm{Jeff}} = \boldsymbol{\Omega}^\infty + \lambda\,\mathbf{p}\times(\mathbf{E}\mathbf{p}), \qquad 
\lambda = \frac{r^2-1}{r^2+1},
\]
where \(\mathbf{p}\) is the symmetry axis, \(\boldsymbol{\Omega}^\infty = \tfrac{1}{2}\boldsymbol{\omega}\) is the background vorticity, and we flip the sign of \(\boldsymbol{\Omega}_{\mathrm{Jeff}}\) to match the convention used by the BEM solver.

\begin{figure}[h!]
    \centering
    \includegraphics[width=0.49\linewidth]{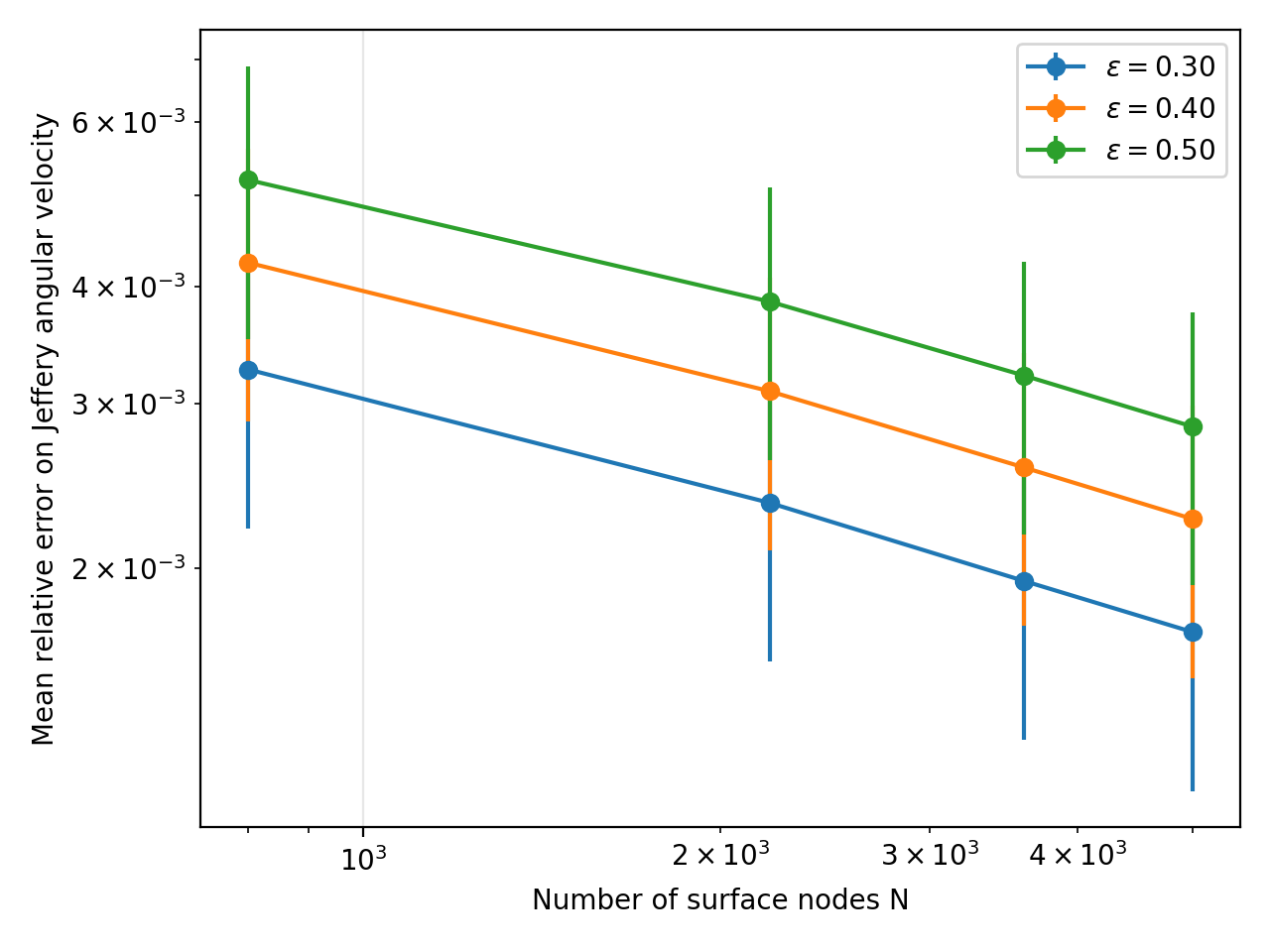}
    \includegraphics[width=0.49\linewidth]{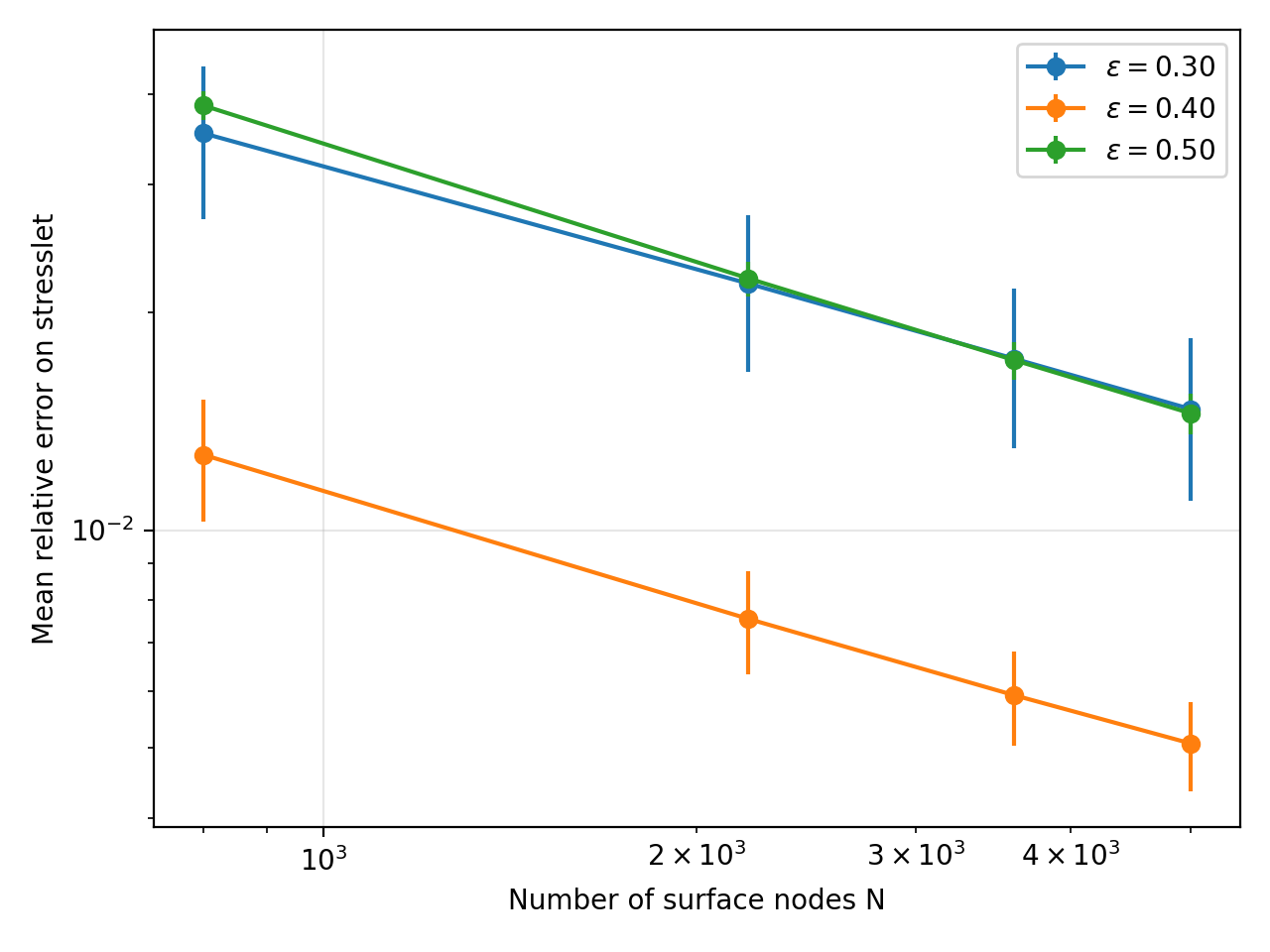}\\
    \includegraphics[width=0.49\linewidth]{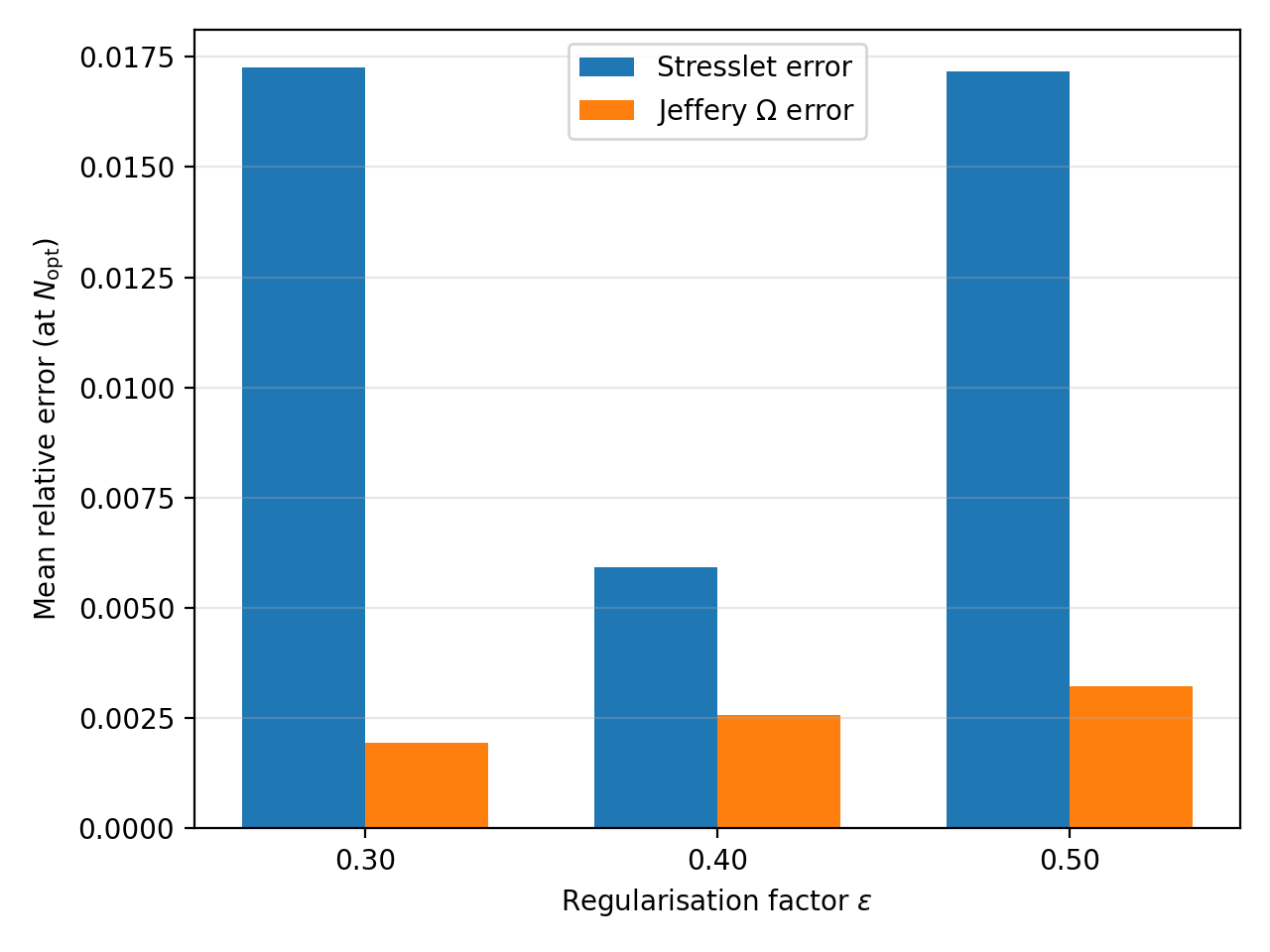}
    \caption{Self-convergence of the spheroidal BEM solver with respect to the number of surface nodes \(N\) and the regularisation factor \(\varepsilon\) for a spheroid with aspect ratio \(r=2\). Top-left: mean relative error in Jeffery angular velocity compared with the analytical solution as a function of \(N\) for three \(\varepsilon\). Top-right: corresponding mean relative error in the deviatoric stresslet. Bottom: mean relative errors at \(\bar{N}=4300\), showing that \(\varepsilon=0.4\) minimises the stresslet error while the angular-velocity error remains small and only weakly dependent on \(\varepsilon\).
}
    \label{fig:BEM_val_spheroids}
\end{figure}

The plots in the top panel of Fig.~\ref{fig:BEM_val_spheroids} show the mean relative errors in Jeffery angular velocity and deviatoric stresslet as functions of the number of surface nodes \(N\) for three regularisation factors \(\varepsilon_{reg} = \varepsilon \sqrt{A_p/N}\) with \(\varepsilon = 0.30, 0.40, 0.50\).
Errors decrease monotonically with \(N\) and all curves approach the same limit; for \(N=3.6\times10^{3}\) and \(\varepsilon=0.40\) we obtain mean errors \(\langle \mathrm{err}(S)\rangle \approx 5.9\times10^{-3}\) and \(\langle \mathrm{err}(\Omega)\rangle \approx 2.6\times10^{-3}\).
The figure in the bottom panel compares the errors at fixed \(N\) and shows that \(\varepsilon=0.40\) minimises the stresslet error while the angular-velocity error depends only weakly on \(\varepsilon\).
Based on these trends we adopt \(\bar{N} = 4.3\times10^{3}\) surface nodes and \(\bar{\varepsilon} = 0.4\) for all spheroidal training data.

As an additional check we vary the aspect ratio \(r=c/a\) at \((N_{\mathrm{opt}},\varepsilon_{\mathrm{opt}})\) and compare with the analytical spheroid solutions. In the spherical limit \(r\to 1\) we recover the known expressions for a rigid sphere, 
\[
S_{ij} = 20\pi\mu a^{3} E_{ij}, 
\qquad 
\boldsymbol{\Omega} = \boldsymbol{\omega},
\]
with mean relative errors \(\langle \mathrm{err}(S_{\mathrm{dev}})\rangle \approx 1.6\times10^{-3}\) and \(\langle \mathrm{err}(\Omega_{\mathrm{Jeff}})\rangle \approx 1.7\times10^{-7}\).

Finally, we verify a set of physics constraints at \((N_{\mathrm{opt}},\varepsilon_{\mathrm{opt}})\).
The computed stresslet is symmetric up to machine precision and the residual net force and torque on the particle are below \(10^{-13}\,\mathrm{N}\) and \(10^{-13}\,\mathrm{N\,m}\), respectively, while the dimensionless spurious translation \(|\mathbf{U}|/(L\|\mathbf{E}\|)\) is of order \(10^{-4}\).
Frame objectivity is satisfied: after applying random rigid rotations to both geometry and flow, the transformed stresslet and angular velocity differ from the rotated reference by at most \(\mathcal O(10^{-5})\) in relative norm, and the translation by at most \(1.3\times10^{-2}\).
A rescaling test, in which the imposed rate is doubled, confirms linearity of the solver, with deviations from \(S(2\dot{\gamma}) = 2S(\dot{\gamma})\) and \(\Omega(2\dot{\gamma}) = 2\Omega(\dot{\gamma})\) at the level of numerical round-off (\(<10^{-15}\)).

\subsubsection*{Validation for helicoidal particles}

For helicoidal particles we consider a slender helical filament with centreline radius \(0.5~\mu\mathrm{m}\), pitch \(2.0~\mu\mathrm{m}\), three turns and wire radius \(0.05~\mu\mathrm{m}\).
The validation dataset combines several particle orientations (uniformly distributed on the unit sphere), five linear flow types (simple shear in the three coordinate planes, uniaxial and biaxial extension) and both right- and left-handed helices.
For each configuration we run the BEM solver for three resolutions \(N=\{2000, 4000, 6000\}\) and three values of the dimensionless regularisation factor \(\varepsilon=\{0.30,0.40,0.50\}\), and collect the resulting stresslets, rigid-body velocities and torques.
From these runs we build ensemble averages and error statistics, which are then used for the convergence, Richardson, chirality, frame-objectivity and linearity tests described below.

\begin{figure}[h!]
    \centering
    \includegraphics[width=0.5\linewidth]{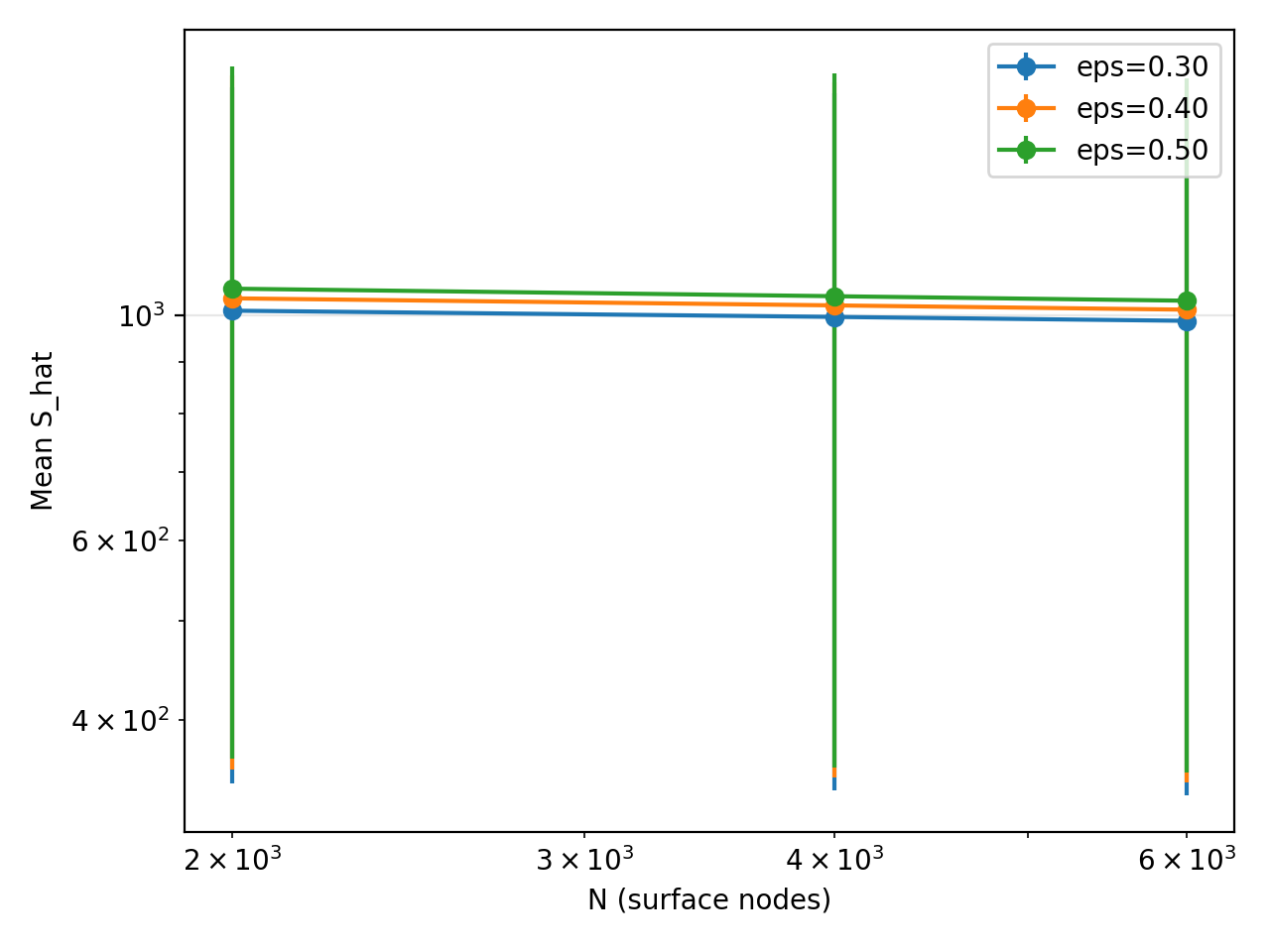}
    \caption{Convergence of the dimensionless stresslet magnitude \(\hat{S}\) for helicoidal particles. The markers show the mean \(\hat{S}\) over the validation dataset as a function of the number of surface nodes \(N\) for three values of the dimensionless regularisation factor \(\varepsilon\) (with \(\varepsilon_{\mathrm{reg}} = \varepsilon\sqrt{A_p/N}\)). Error bars denote one standard deviation over all orientations, flow types and handednesses: they are large because \(\hat{S}\) varies strongly across configurations, not because of numerical noise. The nearly parallel curves and weak dependence on \(N\) and \(\varepsilon\) indicate that the stresslet level is already close to convergence in the range \(N=2000\text{--}6000\).}
    \label{fig:helicoid_conv_Shat}
\end{figure}

We characterise the stresslet through the dimensionless magnitude
\(\hat{S} = \|S_{\mathrm{dev}}\| / (\mu \dot{\gamma} a_{\mathrm{eq}}^{3})\), where \(S_{\mathrm{dev}}\) is the deviatoric part of the stresslet and \(a_{\mathrm{eq}}\) is the radius of a sphere with the same filament volume.
Figure~\ref{fig:helicoid_conv_Shat} shows the mean \(\hat{S}\) as a function of \(N\) for the three \(\varepsilon\).
The curves differ only weakly over the range \(N=2000\text{--}6000\), and the dependence on \(\varepsilon\) is negligible.
For consistency with the spheroid case we therefore fix the dimensionless regularisation factor to \(\varepsilon_{\mathrm{opt}}=0.40\) and adopt a helical resolution \(N_{\mathrm{helix}}=4.3\times10^{3}\).

Due to the absence of analytical solutions, to quantify the discretisation error on the stresslet we perform a Richardson extrapolation using the model 

$$\hat{S}(N) = \hat{S}_\infty + c N^{-p}.$$

\noindent
A global fit over the validation dataset yields \(p\simeq 0.5\) and an estimate of the asymptotic value \(\hat{S}_\infty\).
Per-case fits then provide an empirical distribution of the relative error at \(N_{\mathrm{helix}}\), with mean \(4.3\times10^{-2}\), median \(4.0\times10^{-2}\), 90th–99th percentiles in the range \((5.1\text{--}5.4)\times10^{-2}\), and maximum \(5.4\times10^{-2}\).
This shows that individual helical stresslets used for training carry a discretisation uncertainty of order \(4\text{--}5\%\).

At \((N_{\mathrm{helix}},\varepsilon_{\mathrm{opt}})\) the stresslet remains strictly symmetric and the residual net force and torque are numerically zero (mean norms \(\|\mathbf{F}\|\approx 10^{-22}\,\mathrm{N}\) and \(\|\mathbf{T}\|\approx 10^{-28}\,\mathrm{N\,m}\)).
The dimensionless translation leak \(|\mathbf{U}|/(L\|\mathbf{E}\|)\) is \(\mathcal O(10^{-3})\), while the rotation leak \(|\boldsymbol{\Omega}|/\|\mathbf{E}\|\) is \(\mathcal O(1)\), consistent with a chiral rotor in shear.
Frame objectivity is satisfied to machine precision: after random rigid rotations of both geometry and flow, the transformed stresslet, translation and angular velocity differ from the rotated reference by at most \(10^{-14}\), \(10^{-13}\) and \(10^{-14}\), respectively, in relative norm.
Chirality tests comparing right- and left-handed helices confirm that the axial thrust changes sign while the stresslet and angular velocity remain invariant (thrust antisymmetry ratios \(\lesssim 10^{-11}\); relative variations in stresslet and rotation \(\lesssim 10^{-14}\)).
Finally, a rescaling test with a factor \(\alpha=2\) in the imposed rate yields deviations from exact linearity, \(S(2\dot{\gamma})-2S(\dot{\gamma})\) and \(\Omega(2\dot{\gamma})-2\Omega(\dot{\gamma})\), below \(10^{-15}\), i.e. at numerical round-off level.

%%%%%%%%%%%%%%%%%%%%%%%%%%%%%%%%%%%%%%%%%%%%%%%%%%%%%%%%%%%%%%%%%
\section{Neural Network Surrogate}
\label{sec:Surrogate}

In the previous section we have discussed how the response of a single--particle of fixed shape and orientation to a local linear Stokes flow with rate-of-strain tensor $\mathbf{E}$ and vorticity tensor $\mathbf{W}$ can be written as a linear map of $(\mathbf{E},\mathbf{W})$ through geometry-dependent response tensors.
In particular, the stresslet $\mathbf{S}$ is expressed in terms of the dimensionless fourth-order tensor $\mathsf{M}$ as
\begin{equation}
  \mathbf{S}(\mathbf{E};\mu,\ell,\text{shape},\widehat{\mathbf{p}}) 
  = \mu\,\ell^{3}\,\mathsf{M}(\text{shape},\widehat{\mathbf{p}}):\mathbf{E},
\end{equation}
where $\mu$ is the dynamic viscosity, $\ell$ is a characteristic particle length, and $\widehat{\mathbf{p}}$ is the unit vector along the particle axis.
Analogous third-order tensors $\mathsf{M}^{U}$ and $\mathsf{M}^{\Omega}$ determine the chiral thrust $\mathbf{U}$ (for helices) and the rigid-body angular velocity $\boldsymbol{\Omega}$, respectively.

Rather than tabulating $\mathsf{M}$, $\mathsf{M}^{U}$ and $\mathsf{M}^{\Omega}$ explicitly for all orientations and shapes, we seek a data-driven closure that directly approximates the corresponding input–output map.
Symbolically, for a particle of given geometry we aim to learn
\begin{equation}
  (\mathbf{E},\mathbf{W},\widehat{\mathbf{p}},h)
  \;\longmapsto\;
  (\mathbf{S},\mathbf{U},\boldsymbol{\Omega}),
\end{equation}
where $h=\pm 1$ encodes the handedness for helicoidal particles and is absent in the spheroidal case (for which $\mathbf{U}\equiv\mathbf{0}$).
In practice, the neural surrogate takes as input the Cartesian components of $\mathbf{E}$ and $\mathbf{W}$ together with the particle axis (and, for helices, $h$), and returns the corresponding components of the stresslet, chiral thrust and rigid-body rotation as computed by the BEM solver.
Training is performed at fixed $\mu=1$, unit characteristic rate and fixed reference length for each shape, so that the network effectively learns the dimensionless response encoded in the tensors $\mathsf{M}$, $\mathsf{M}^{U}$ and $\mathsf{M}^{\Omega}$, while the trivial rescalings discussed in Sec.~\ref{ssec:linearity} with viscosity, length and rate are reinstated analytically in the FCM solver.

\paragraph{Training dataset}

To train the neural surrogate we first generated supervised datasets using the validated BEM solvers for both geometries (prolate spheroids and helicoidal particles). Each sample corresponds to a rigid particle immersed in a canonical linear Stokes flow with prescribed rate-of-strain tensor $\mathbf{E}$, vorticity tensor $\mathbf{W}$, and particle axis direction $\mathbf{p}$; the BEM solver returns the corresponding hydrodynamic stresslet and rigid-body rotation (and, for helices, the chiral thrust).

For the spheroidal reference case we considered a prolate spheroid of aspect ratio
\[
r = \frac{c}{a} = 2,
\]
with semi-axes $a = 1$ and $c = 2$ in nondimensional units. The local flow was restricted to four canonical incompressible linear flows: uniaxial extension, planar extension, biaxial extension, and simple shear. For each flow, 256 orientations of the spheroid symmetry axis were sampled on the unit sphere using a Fibonacci lattice, yielding a uniform distribution in orientation space. This gives a dataset of $4 \times 256 = 1024$ BEM evaluations at fixed viscosity $\mu = 1$ and rate $\dot{\gamma} = 1$, with a surface discretization of $N = 2500$ nodes for all cases. The input vector for each sample is
\[
\mathbf{X}
=
\bigl(
E_{xx}, E_{yy}, E_{zz},
E_{xy}, E_{xz}, E_{yz},
W_{xy}, W_{xz}, W_{yz},
p_x, p_y, p_z
\bigr)
\in \mathbb{R}^{12},
\]
i.e. the independent components of $\mathbf{E}$ and $\mathbf{W}$ plus the three components of the spheroid axis. The BEM solver returns the $3 \times 3$ stresslet and the rigid-body angular velocity, which we store as
\[
\mathbf{Y}
=
\bigl(
S_{xx}, S_{yy}, S_{zz},
S_{xy}, S_{xz}, S_{yz},
\Omega_x, \Omega_y, \Omega_z
\bigr)
\in \mathbb{R}^{9}.
\]
All samples from this BEM-based dataset are then split into training, validation, and test sets using a stratified shuffle procedure (80/10/10) that preserves the relative frequency of the four flow types in each subset.

The helicoidal dataset is constructed analogously, but for a fixed three-turn helix of prescribed radius, pitch, and wire thickness. The geometry is kept constant across all samples (again exploiting linearity in Stokes flow with respect to length and rate), while we vary flow type, orientation, and chirality. We use the same four canonical linear flows and the same Fibonacci sampling with 256 orientations of the helix axis. In addition, we explicitly sample both right-handed and left-handed helices, encoded by a handedness parameter $h = \pm 1$. With 4 flows, 256 orientations, and 2 chiralities this yields $4 \times 256 \times 2 = 2048$ BEM cases. The helical surface is discretized with $N = 4300$ nodes, and the regularization factor $\varepsilon=0.4$

The input feature vector extends the spheroidal one by the handedness,
\[
\mathbf{X}
=
\bigl(
E_{xx}, E_{yy}, E_{zz},
E_{xy}, E_{xz}, E_{yz},
W_{xy}, W_{xz}, W_{yz},
p_x, p_y, p_z, h
\bigr)
\in \mathbb{R}^{13},
\]
while the output includes the $3 \times 3$ stresslet, the chiral thrust vector $\mathbf{U}$, and the rigid-body angular velocity,
\[
\mathbf{Y}
=
\bigl(
S_{xx}, S_{yy}, S_{zz},
S_{xy}, S_{xz}, S_{yz},
U_x, U_y, U_z,
\Omega_x, \Omega_y, \Omega_z
\bigr)
\in \mathbb{R}^{12}.
\]
This dataset is again partitioned into training, validation, and test sets with the same 80/10/10 stratified strategy, ensuring all flows and both chiralities are represented in each split.

\paragraph{Training}
For both particle geometries we approximate the mapping 
$$(\mathbf{E},\mathbf{W},\widehat{\mathbf{p}},h)\mapsto(\mathbf{S},\mathbf{U},\boldsymbol{\Omega})$$ 
with a fully connected feed–forward neural network (FCNN).  
In all results reported below we use the “large’’ architecture, consisting of four hidden layers with widths $(256,\,256,\,128,\,64)$, followed by a linear output layer.
All hidden layers use a \texttt{tanh} activation, so that the surrogate is a smooth nonlinear map of the input features.
For the spheroidal surrogate the input layer can have dimension up to \(d_{\text{in}} = 21\), corresponding to the 12 raw features \((\mathbf{E},\mathbf{W},\widehat{\mathbf{p}})\) possibly augmented by 9 physics–motivated scalars built from \(\mathbf{E}\) and \(\widehat{\mathbf{p}}\)
(see below), and the output layer has \(d_{\text{out}} = 9\) components (symmetric stresslet and angular velocity).
For helicoidal particles we use the same hidden architecture and feature set, but add the handedness parameter \(h=\pm 1\) as an extra input, so that \(d_{\text{in}} = 22\), and we can option to extend the output to \(d_{\text{out}} = 12\) components to include the chiral thrust vector.
This choice allows us to treat spheroids and helices within a common FCNN architecture, with only the input and output dimensions changed.

\begin{figure}[h!]
    \centering
    \includegraphics[width=0.49\linewidth]{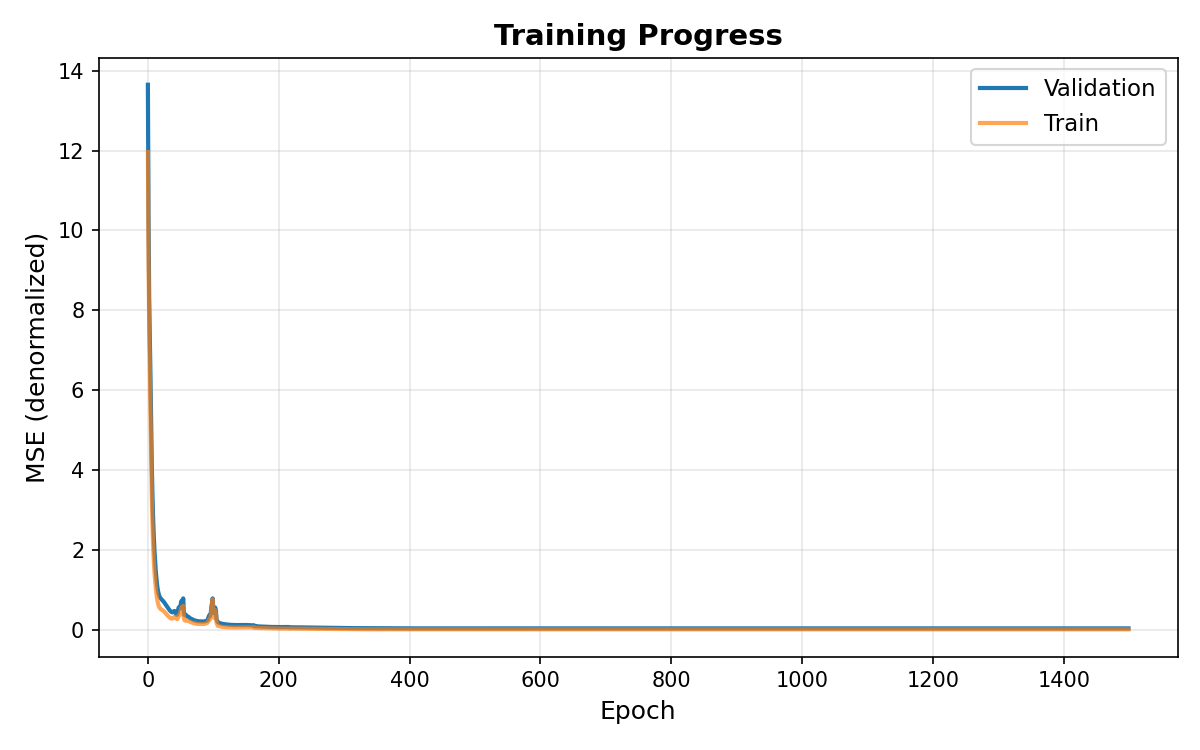}
    \includegraphics[width=0.49\linewidth]{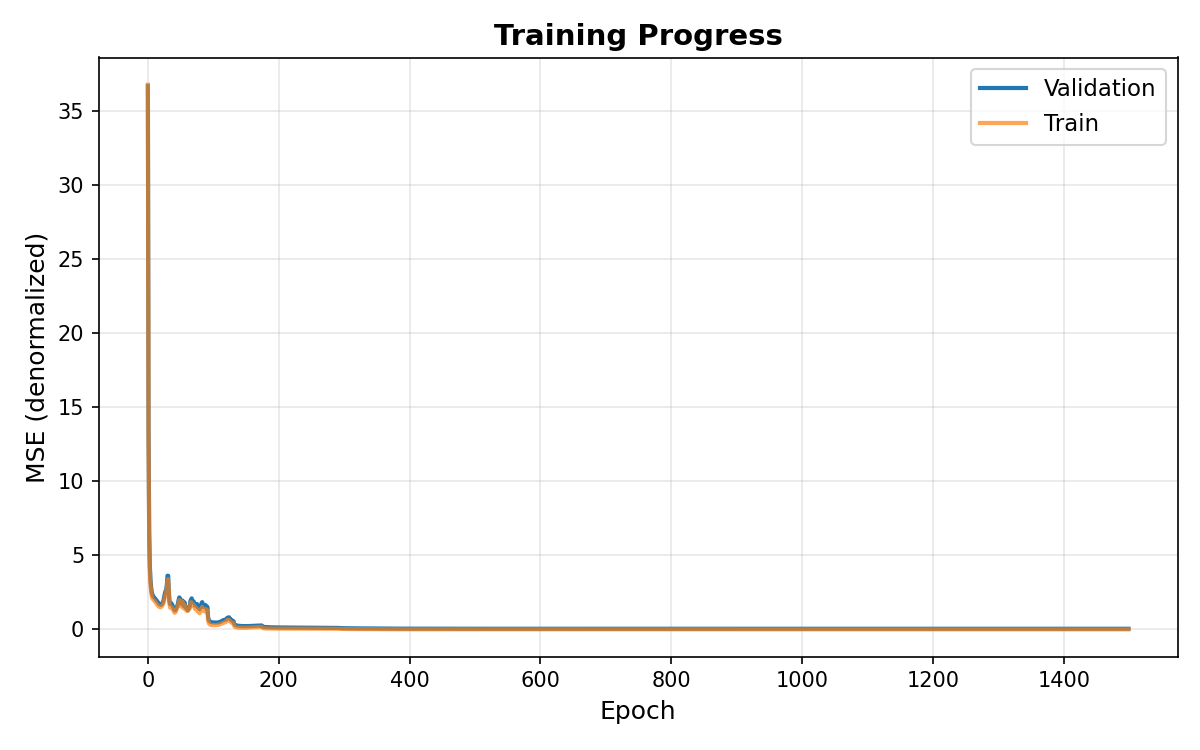}
    \caption{Training history of the large feature–augmented FCNN surrogates for spheroids (left) and helicoidal particles (right). The denormalised training and validation MSE rapidly decrease and then plateau, indicating stable convergence without overfitting.}
    \label{fig:losses}
\end{figure}

Before training, each input and output component is normalised to zero mean and unit variance using the statistics of the training set; the same affine
transformation is then applied to the validation and test sets.
The networks are trained with mini–batch gradient descent using the Adam optimiser with a fixed initial learning rate \(\eta_0 = 10^{-3}\), a mean–squared–error (MSE) loss on the normalised outputs, and a batch size of 32.
We train for at most \(N_{\text{epoch}} = 1500\) epochs, with an adaptive learning–rate schedule (\texttt{ReduceLROnPlateau} on the validation loss) and early stopping with restoration of the best validation weights (losses are plotted in Fig.~\ref{fig:losses}). Random seeds are fixed for NumPy, Python and TensorFlow to ensure reproducible splits and weight initialisation.
All models are implemented in Python using TensorFlow/Keras, and the same training pipeline (normalisation, optimiser, and callbacks) is used for both geometries~\cite{laudato2025_neuralfcm_github}.

In addition to this reference “large’’ network with engineered features, we also trained smaller “moderate’’ architectures and variants without feature
augmentation.
These ablations are used to assess the impact of network capacity and feature engineering on accuracy, and are discussed in detail in the following section.

\begin{figure}[h!]
    \centering
    \includegraphics[width=\linewidth]{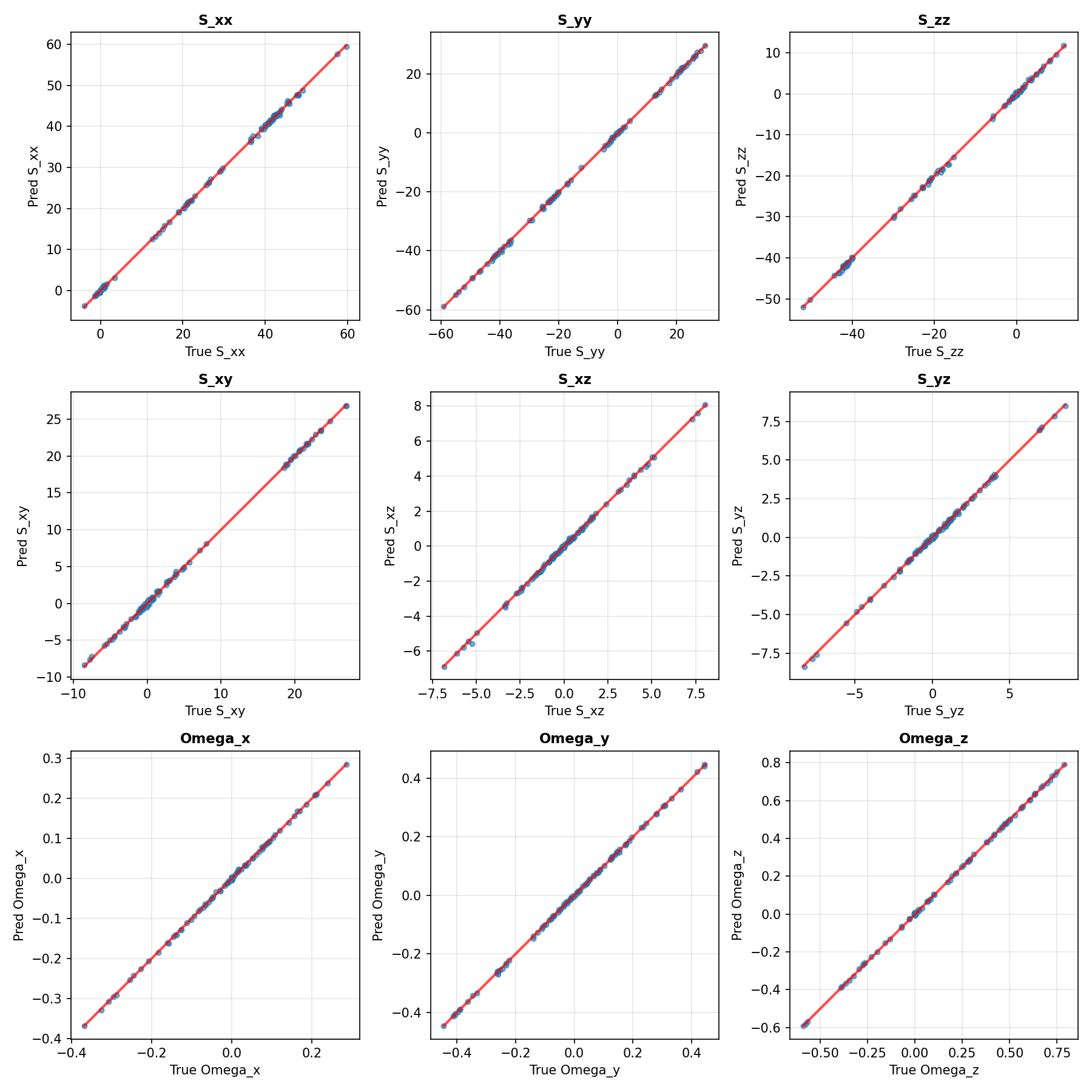}
    \caption{Parity plots comparing FCNN surrogate predictions with BEM reference values for the best spheroidal model on the test set.}
    \label{fig:parity_spheroids}
\end{figure}

\begin{figure}[h!]
    \centering
    \includegraphics[width=\linewidth]{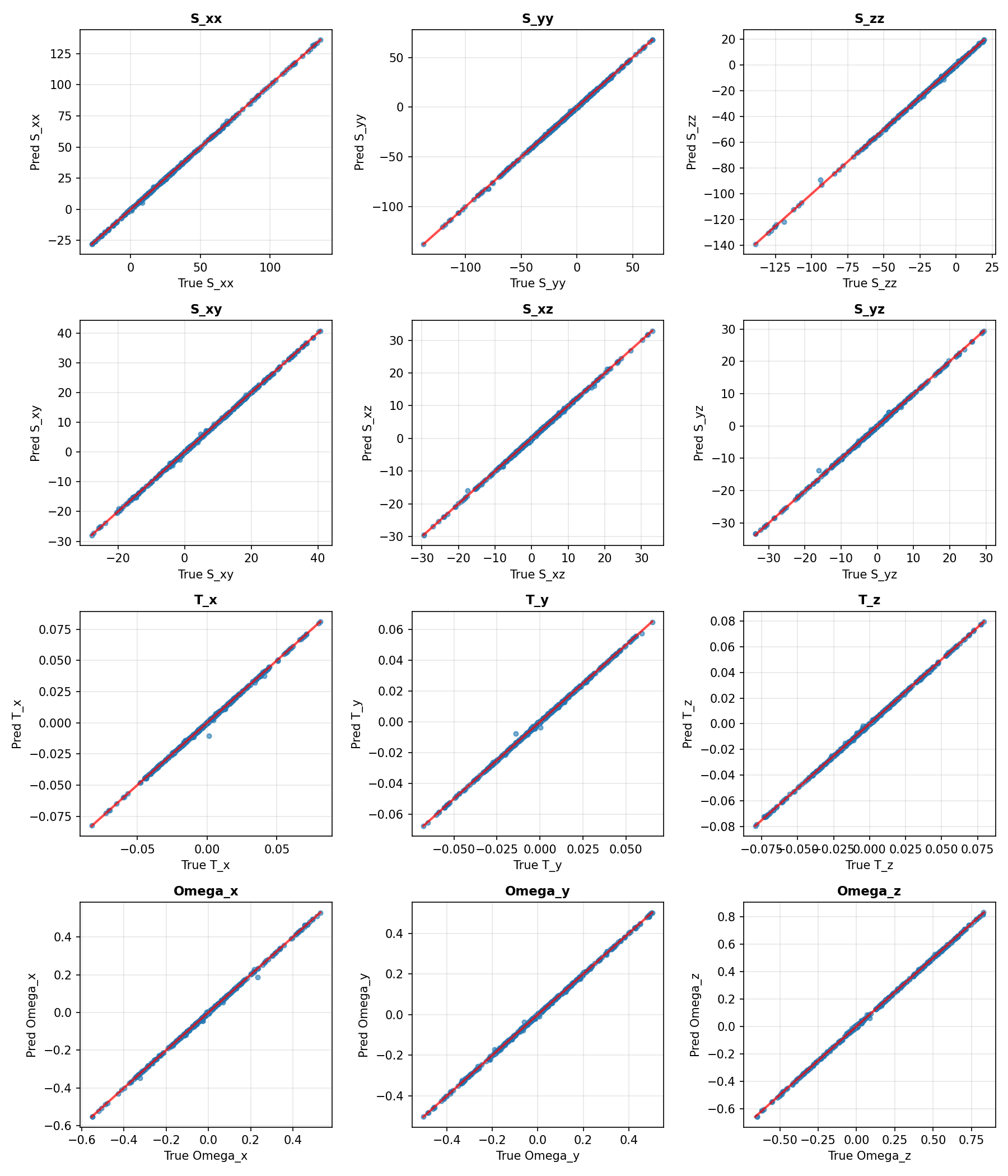}
    \caption{Parity plots comparing FCNN surrogate predictions with BEM reference values for the best helicoidal particles model on the test set.}
    \label{fig:parity_helicoidals}
\end{figure}

\paragraph{Overall accuracy}
For each geometry we select as reference the best-performing surrogate among the architectures explored in the ablation study, namely the large network with engineered features (and, for helices, handedness augmentation).
On the held-out test set, the spheroidal surrogate (with features, large network) achieves a median relative error in the deviatoric stresslet magnitude of $0.87\%$, with a $95$th percentile of $2.06\%$, and a median relative $\ell_2$ error in the angular velocity $\boldsymbol{\Omega}$ of $1.29\%$.  
For the helicoidal surrogate (with features, large network, and handedness augmentation) the corresponding median relative errors are $0.70\%$ for the deviatoric stresslet, $0.79\%$ for $\boldsymbol{\Omega}$, and $0.89\%$ for the chiral thrust $\mathbf{U}$, with a $95$th percentile error of $2.96\%$ for the deviatoric stresslet.  
These values indicate that, for both geometries, the FCNN surrogate reproduces the BEM input–output map to within $\mathcal{O}(1\%)$ median relative error, with only a small fraction of samples exceeding $2$--$3\%$ (see Figs.~\ref{fig:parity_spheroids} and \ref{fig:parity_helicoidals}).
A more detailed per-component analysis of the error, including the role of low-energy components and the resulting uncertainty in rheological observables, is discussed in the next section.

\section{Surrogate Accuracy and Error Propagation in FCM}
\label{sec:accuracy}
\subsection{Surrogate accuracy: effect of input features and network size}
\label{subsec:ablation}

To assess the robustness of the surrogate, we performed a systematic ablation study in which we varied (i) the input representation and (ii) the size of the feed–forward network, while keeping all other training hyperparameters fixed. In all cases the input encodes the local linearised flow and particle orientation,
\[
\mathbf{x}_{\mathrm{raw}} = \big(E_{xx},E_{yy},E_{zz},E_{xy},E_{xz},E_{yz},
W_{xy},W_{xz},W_{yz},
p_x,p_y,p_z\big)\in\mathbb{R}^{12},
\]
where $E$ is the symmetric rate–of–strain tensor, $W$ is the antisymmetric spin tensor, and $\mathbf{p}$ is the particle orientation.

\paragraph{Physics–informed feature engineering}
In the \emph{features} configurations we augment $\mathbf{x}_{\mathrm{raw}}$ with nine additional physics–informed scalars that encode the tensorial structures that appear in classical suspension theory:
\begin{itemize}
\item strain acting on the particle axis $\Rightarrow\mathbf{E}\mathbf{p}\in\mathbb{R}^3$,
\item extensional or compressive strain along $\mathbf{p}$
$\Rightarrow \mathbf{p}^{\mathsf{T}}\mathbf{E}\mathbf{p}\in\mathbb{R}$,
\item Frobenius norms of strain and spin $\Rightarrow\|\mathbf{E}\|_F,\ \|\mathbf{W}\|_F $,
\item Jeffery–like rotation term $\Rightarrow \mathbf{p}\times(\mathbf{E}\mathbf{p})\in\mathbb{R}^3$.
\end{itemize}
The final input is therefore
\[
\mathbf{x}_{\mathrm{eng}} = \big(\mathbf{x}_{\mathrm{raw}},\; \mathbf{E}\mathbf{p},\; \mathbf{p}^{\mathsf{T}}\mathbf{E}\mathbf{p},\;
\|\mathbf{E}\|_F,\;\|\mathbf{W}\|_F,\; \mathbf{p}\times(\mathbf{E}\mathbf{p})\big)\in\mathbb{R}^{21}.
\]
The first group of features, $\mathbf{E}\mathbf{p}$ and $\mathbf{p}^{\mathsf{T}}\mathbf{E}\mathbf{p}$, captures the coupling between the orientation and the local strain and corresponds to the tensorial building blocks that appear in resistance–tensor closures. The scalar norms $\|\mathbf{E}\|_F$ and $\|\mathbf{W}\|_F$ encode the overall flow intensity and help the network normalise its response across the range of deformation and vorticity rates. Finally, the vector $\mathbf{p}\times(\mathbf{E}\mathbf{p})$ is directly related to Jeffery’s angular velocity and exposes the dominant structure of the stresslet–rotation coupling to the surrogate.

\paragraph{Network architectures}
We compare two fully–connected architectures:
\begin{itemize}
  \item a \emph{moderate} network with three hidden layers of size $128$–$128$–$64$;
  \item a \emph{large} network with four hidden layers of size $256$–$256$–$128$–$64$.
\end{itemize}
In all cases we use $\tanh$ activations, a small $L^2$ kernel regularisation, and the same optimiser and learning–rate schedule. The output layer has nine components and collects the six independent components of the deviatoric stresslet and the three components of the angular velocity (for helicoidal particles we also monitor the scalar thrust). All errors reported below are relative errors on the held–out test set, expressed as percentages.

\paragraph{Spheroidal particles}
Table~\ref{tab:ablation_spheroids} summarises the ablation study for spheroids. Starting from the baseline model that receives only the raw $(E,W,\mathbf{p})$ input, adding more parameters (``no features + large'') produces only a modest improvement, with the median deviatoric–stresslet error decreasing from $9.44\%$ to $8.53\%$ and the median angular–velocity error remaining of order $15\%$. In contrast, introducing the physics–informed features at fixed architecture yields a much larger gain. For the moderate network, the median error on the deviatoric stresslet decreases from $9.44\%$ to $6.38\%$, and the median angular–velocity error from $13.71\%$ to $8.48\%$. Removing regularization and tuning the training seed leads to a best configuration with median errors as low as $0.75\%$ for the deviatoric stresslet and $1.18\%$ for the angular velocity, while keeping the same moderate architecture. This shows that even low regularization levels induce bias errors in the model and that the physics–informed feature set is a dominant source of improvement over brute–force increases in network capacity .

\begin{table}
  \centering
  \small
  \begin{tabular}{lccc}
    \hline
    Configuration & $\mathrm{median} \,|\Delta S_{\mathrm{dev}}|$ & $95\mathrm{th}\,|\Delta S_{\mathrm{dev}}|$ & $\mathrm{median}\,|\Delta\boldsymbol{\Omega}|$ \\
    \hline
    No features, moderate & $9.44$ & $16.79$ & $13.71$ \\
    No features, large    & $8.53$ & $17.03$ & $16.35$ \\
    Features, moderate    & $6.38$ & $12.97$ & $8.48$  \\
    Features, large       & $7.06$ & $13.48$ & $11.41$ \\
    Features, moderate, low reg. & $5.58$ & $10.07$ & $6.95$ \\
    Features, large, low reg.    & $5.14$ & $8.86$  & $8.93$ \\
    Features, moderate, no reg.  & $0.87$ & $2.06$  & $1.29$ \\
    Features, large, no reg. & $0.75$ & $1.71$ & $1.18$ \\
    \hline
  \end{tabular}
  \caption{Ablation study for the spheroidal surrogate. The table reports median and $95$th–percentile relative errors (in percent) on the Frobenius norm of the deviatoric stresslet, together with the median relative error on the angular–velocity norm, for different combinations of input features and network size. All models are trained on the same dataset and evaluated on the same test set.}
  \label{tab:ablation_spheroids}
\end{table}

\paragraph{Helicoidal particles and chiral augmentation}
For helicoidal particles the mapping from $(\mathbf{E},\mathbf{W},\mathbf{p})$ to deviatoric stresslet, angular velocity, and thrust is more nonlinear and strongly chiral than in the spheroidal case.
The ablation study (Table~\ref{tab:ablation_helicoids}) shows that three ingredients play distinct roles: feature engineering, $L^2$ regularisation, and chiral data augmentation.
Starting from the baseline, moderately sized network with raw inputs and standard regularisation, the median error on the deviatoric stresslet is $18.51\%$. Adding the physics-informed features reduces this to $10.22\%$, roughly a factor of two improvement, while changing the network size from moderate to large at fixed regularisation has only a marginal effect.

The most dramatic gain comes from removing the $L^2$ regularisation.
Even without any additional features, the median deviatoric-stresslet error drops from $18.51\%$ to $2.94\%$, and the angular velocity and thrust errors fall to a few percent.
Introducing the features on top of the unregularised baseline yields a further but comparatively modest refinement, from $2.94\%$ to $2.82\%$ median stresslet error, with similarly small changes for the other quantities; increasing the network size again has only a minor impact.
Finally, augmenting the training set with chiral pairs (mirror configurations with opposite helicity) is what pushes the surrogate into the sub-percent regime. With chiral augmentation, the median stresslet error decreases to $0.87\%$ for the moderate network and $0.70\%$ for the large one, while the median errors on angular velocity and thrust remain below $1\%$.
Overall, for helicoidal particles the dominant effects are the removal of over-regularisation and the enforcement of chiral symmetry through data augmentation, whereas increasing network capacity and adding features play a secondary, fine-tuning role.

\begin{table}
  \centering
  \small
  \setlength{\tabcolsep}{3pt} % default is 6pt
\begin{tabular}{lcccc}
    \hline
    Configuration & $\mathrm{median} \,|\Delta S_{\mathrm{dev}}|$ & $95\mathrm{th}\,|\Delta S_{\mathrm{dev}}|$ & $\mathrm{median}\,|\Delta\boldsymbol{\Omega}|$
           & $\mathrm{median}\,\Delta\boldsymbol{T}$ \\
    \hline
    B–M–R            & 18.51 & 34.13 & 16.35 & 21.52 \\
    F–M–R            & 10.22 & 20.36 &  9.35 & 11.11 \\
    F–L–R            & 10.50 & 18.63 & 10.13 & 10.39 \\
    B–M–NR           &  2.94 &  7.32 &  2.86 &  4.12 \\
    F–M–NR           &  2.82 &  7.50 &  2.44 &  3.38 \\
    F–L–NR           &  2.70 &  6.57 &  2.36 &  3.67 \\
    B–M–NR–Ch        &  1.18 &  3.44 &  1.32 &  1.53 \\
    B–L–NR–Ch        &  0.99 &  2.92 &  1.06 &  1.22 \\
    F–M–NR–Ch        &  0.87 &  2.87 &  1.01 &  1.16 \\
    F–L–NR–Ch        &  0.70 &  2.96 &  0.79 &  0.89 \\
    \hline
  \end{tabular}
    \caption{Ablation study for the helicoidal surrogate. Configurations are abbreviated as: B = baseline input, F = with features,  M = moderate net, L = large net, R = with $L^2$ regularisation, NR = no regularisation, Ch = chiral data augmentation. Errors are relative and expressed in percent.}
  \label{tab:ablation_helicoids}
\end{table}

Overall, the ablation results indicate that the physics–informed feature set captures most of the relevant structure of the microhydrodynamic closure. Once these features are provided, even a moderate network attains sub–percent errors on all quantities of interest, and additional increases in network size yield only marginal gains compared with the effect of feature engineering and chiral augmentation.

\subsection{Error decomposition for the stresslet surrogate}
\label{ssec:stresslet_error}

In this subsection we quantify the total error of the surrogate at the level of the deviatoric stresslet.
In the FCM formulation only the stresslet enters the particle contribution to the bulk stress, while the rotation and (for helicoidal particles) the thrust affect only the kinematics.
We therefore restrict the error analysis to the deviatoric stresslet and use it as the reference quantity for uncertainty quantification. We distinguish three stresslets:
\begin{itemize}
  \item ideal, infinite--resolution Stokes solution: $\mathbf{S}_\infty$,
  \item BEM solution at chosen $N,\varepsilon$: $\mathbf{S}_{\text{BEM}}$,
  \item surrogate prediction: $\mathbf{S}_{\text{NN}}.$
\end{itemize}
They are related by
\begin{equation}
  \mathbf{S}_{\text{BEM}} = \mathbf{S}_\infty + \delta \mathbf{S}_{\text{BEM}},
  \qquad
  \mathbf{S}_{\text{NN}} = \mathbf{S}_{\text{BEM}} + \delta \mathbf{S}_{\text{NN}},
\end{equation}
so that the total error with respect to the ideal solution reads
\begin{equation}
  \mathbf{S}_{\text{NN}} - \mathbf{S}_\infty
  = \delta \mathbf{S}_{\text{BEM}} + \delta \mathbf{S}_{\text{NN}}.
\end{equation}
We measure errors using the Frobenius norm of the deviatoric stresslet and define the relative errors
\begin{equation}
  e_{\text{BEM}} = \frac{\|\delta \mathbf{S}_{\text{BEM}}\|_F}{\|\mathbf{S}_\infty\|_F},
  \qquad
  e_{\text{NN}} = \frac{\|\delta \mathbf{S}_{\text{NN}}\|_F}{\|\mathbf{S}_{\text{BEM}}\|_F},
  \qquad
  e_{\text{tot}} = \frac{\|\mathbf{S}_{\text{NN}} - \mathbf{S}_\infty\|_F}{\|\mathbf{S}_\infty\|_F}.
\end{equation}
Since both $e_{\text{BEM}}$ and $e_{\text{NN}}$ are small, we approximate
$\|\mathbf{S}_\infty\|_F \approx \|\mathbf{S}_{\text{BEM}}\|_F$ and express all errors as percentages of the same typical magnitude. Under this approximation the total relative error satisfies
\begin{equation}
  e_{\text{tot}} \;\lesssim\; e_{\text{BEM}} + e_{\text{NN}}
  \label{eq:worst_case_sum}
\end{equation}
in a worst--case scenario where the two contributions are systematic in the same direction (same sign), and
\begin{equation}
  e_{\text{tot}} \;\approx\; \sqrt{e_{\text{BEM}}^2 + e_{\text{NN}}^2}
  \label{eq:rms_sum}
\end{equation}
if they are treated as independent and uncorrelated error sources.

\paragraph{Spheroidal particles}
For spheroids the BEM solver is benchmarked directly against analytical solutions. At the resolution used to generate the training and test sets, the BEM stresslet error is
\begin{equation}
  e_{\text{BEM}}^{\text{sph}} \approx 0.6\% .
\end{equation}
The final surrogate for spheroids attains a median relative error of
\begin{equation}
  e_{\text{NN,\,med}}^{\text{sph}} \approx 0.75\%, 
  \qquad
  e_{\text{NN,\,95}}^{\text{sph}} \approx 1.71\% 
\end{equation}
on the held--out test set (Frobenius norm of the deviatoric stresslet).

Combining BEM and surrogate errors via \eqref{eq:worst_case_sum} yields conservative bounds
\begin{equation}
  e_{\text{tot,\,med}}^{\text{sph}} \lesssim 0.6\% + 0.75\% \approx 1.35\%,
  \qquad
  e_{\text{tot,\,95}}^{\text{sph}} \lesssim 0.6\% + 1.71\% \approx 2.31\%.
\end{equation}
If we instead use the RMS estimate \eqref{eq:rms_sum}, we obtain
\begin{equation}
  e_{\text{tot,\,med}}^{\text{sph}} \approx 1.0\%,
  \qquad
  e_{\text{tot,\,95}}^{\text{sph}} \approx 1.8\%,
\end{equation}
which shows that, for spheroids, the combined closure error of the BEM--surrogate chain remains at the level of one to two percent with respect to the ideal Stokes solution.

\paragraph{Helicoidal particles}
For helicoidal particles no analytical solution is available. Instead, the BEM accuracy is quantified via a Richardson extrapolation in the number of surface elements, which gives
\begin{equation}
  e_{\text{BEM,\,med}}^{\text{hel}} \approx 4.0\%,
  \qquad
  e_{\text{BEM,\,95}}^{\text{hel}} \approx 5.1\%
\end{equation}
for the chosen resolution. The final surrogate model (features, large network, no regularisation, chiral augmentation) achieves
\begin{equation}
  e_{\text{NN,\,med}}^{\text{hel}} \approx 0.7\%,
  \qquad
  e_{\text{NN,\,95}}^{\text{hel}} \approx 2.96\%.
\end{equation}
A worst--case aligned--bias estimate \eqref{eq:worst_case_sum} then gives
\begin{equation}
  e_{\text{tot,\,med}}^{\text{hel}} \lesssim 4.0\% + 0.7\% \approx 4.7\%,
  \qquad
  e_{\text{tot,\,95}}^{\text{hel}} \lesssim 5.1\% + 2.96\% \approx 8.1\%,
\end{equation}
while the RMS combination \eqref{eq:rms_sum} yields
\begin{equation}
  e_{\text{tot,\,med}}^{\text{hel}} \approx 4.1\%,
  \qquad
  e_{\text{tot,\,95}}^{\text{hel}} \approx 5.9\%.
\end{equation}
In other words, for helicoidal particles the total stresslet error of the closure is dominated by the BEM discretisation error: the surrogate introduces only a $\mathcal{O}(1\%)$ contribution on top of an intrinsic BEM uncertainty of about $4\text{--}5\%$.

These estimates provide the stresslet--level error budget of the BEM--surrogate pipeline. In the next subsection we will use the same error decomposition to propagate these uncertainties to the particle contribution to the macroscopic stress in FCM.

\subsection{Error propagation to macroscopic rheology}
\label{subsec:rheology_error}

In the FCM formulation the surrogate enters the macroscopic equations only through the particle contribution to the bulk stress. For a suspension of $N_p$ particles in a volume $V$ we write
\begin{equation}
  \boldsymbol{\sigma}_p
  = \frac{1}{V}\sum_{n=1}^{N_p} \mathbf{S}_{\text{dev}}^{(n)},
\end{equation}
where $\mathbf{S}_{\text{dev}}^{(n)}$ is the deviatoric stresslet of particle $n$. Rotation and (for helicoidal particles) thrust affect the particle kinematics but do not enter $\boldsymbol{\sigma}_p$, so we restrict the rheology error analysis to the deviatoric stresslet.

Let $\mathbf{S}_\infty^{(n)}$ denote the ideal stresslet (infinite-resolution Stokes solution) and $\mathbf{S}_{\text{cl}}^{(n)}$ the stresslet used in FCM, obtained from the complete BEM--surrogate closure (BEM + neural network) at the chosen resolution. We write
\begin{equation}
  \mathbf{S}_{\text{cl}}^{(n)} 
  = \mathbf{S}_\infty^{(n)} + \delta \mathbf{S}^{(n)},
\end{equation}
and define the per–particle relative closure error
\begin{equation}
  \varepsilon_{\text{cl}}^{(n)} 
  = \frac{\|\delta \mathbf{S}^{(n)}\|_F}{\|\mathbf{S}_\infty^{(n)}\|_F}.
\end{equation}
From the previous subsection we obtained the following characteristic levels for the total closure error (including both BEM discretisation and surrogate approximation):
\begin{align}
  e_{\text{cl,\,med}}^{\text{sph}} &\approx 1\%, 
  & e_{\text{cl,\,95}}^{\text{sph}} &\approx 2\%, \\
  e_{\text{cl,\,med}}^{\text{hel}} &\approx 4\text{--}5\%,
  & e_{\text{cl,\,95}}^{\text{hel}} &\approx 6\text{--}8\%.
\end{align}
Here $e_{\text{cl,\,med}}$ and $e_{\text{cl,\,95}}$ denote the median and $95$th--percentile of $\varepsilon_{\text{cl}}^{(n)}$ on the test set.

The true particle stress and its closure approximation are
\begin{equation}
  \boldsymbol{\sigma}_p^\infty
    = \frac{1}{V}\sum_{n=1}^{N_p} \mathbf{S}_\infty^{(n)}, 
  \qquad
  \boldsymbol{\sigma}_p^{\text{cl}}
    = \frac{1}{V}\sum_{n=1}^{N_p} \mathbf{S}_{\text{cl}}^{(n)}
    = \boldsymbol{\sigma}_p^\infty 
      + \frac{1}{V}\sum_{n=1}^{N_p} \delta \mathbf{S}^{(n)}.
\end{equation}
We denote the error in the particle stress by
\begin{equation}
  \Delta \boldsymbol{\sigma}_p 
  = \boldsymbol{\sigma}_p^{\text{cl}} - \boldsymbol{\sigma}_p^\infty
  = \frac{1}{V}\sum_{n=1}^{N_p} \delta \mathbf{S}^{(n)},
\end{equation}
and define the relative rheology error
\begin{equation}
  e_{\sigma} 
  = \frac{\|\Delta \boldsymbol{\sigma}_p\|}{\|\boldsymbol{\sigma}_p^\infty\|}.
\end{equation}
To estimate $e_{\sigma}$ in a simple and transparent way, we consider two idealised scenarios for the closure error: a biased (systematic) case and an uncorrelated (random) case.

\paragraph{Biased (systematic) closure error}
In the biased scenario we assume that the closure behaves as an approximately uniform multiplicative perturbation of the ideal stresslet,
\begin{equation}
  \mathbf{S}_{\text{cl}}^{(n)} \approx (1+\epsilon_{\text{cl}})\,\mathbf{S}_\infty^{(n)},
  \qquad
  |\epsilon_{\text{cl}}| \sim e_{\text{cl}},
\end{equation}
where $e_{\text{cl}}$ is a representative closure error level (for instance, the median or $95$th--percentile value from the test set). Inserting this into the definition of $\boldsymbol{\sigma}_p^{\text{cl}}$ yields
\begin{equation}
  \boldsymbol{\sigma}_p^{\text{cl}}
  \approx \frac{1}{V}\sum_{n=1}^{N_p} (1+\epsilon_{\text{cl}})\,\mathbf{S}_\infty^{(n)}
  = (1+\epsilon_{\text{cl}})\,\boldsymbol{\sigma}_p^\infty,
\end{equation}
and therefore
\begin{equation}
  e_{\sigma}^{\text{bias}} 
  \equiv \frac{\|\Delta \boldsymbol{\sigma}_p\|}{\|\boldsymbol{\sigma}_p^\infty\|}
  \approx |\epsilon_{\text{cl}}| \approx e_{\text{cl}}.
\end{equation}
Under this conservative assumption the rheology error directly mirrors the closure error. Using the combined BEM-surrogate error levels from above we obtain
\begin{align}
  e_{\sigma,\text{med}}^{\text{bias,\,sph}} &\approx 1\%,
  & e_{\sigma,\text{95}}^{\text{bias,\,sph}} &\approx 2\%, \\
  e_{\sigma,\text{med}}^{\text{bias,\,hel}} &\approx 4\text{--}5\%,
  & e_{\sigma,\text{95}}^{\text{bias,\,hel}} &\approx 6\text{--}8\%.
\end{align}
In other words, even in a worst--case scenario where the closure induces a fully correlated bias across all particles, the resulting error in the particle contribution to the bulk stress remains at the level of a few percent: about $1\text{--}2\%$ for spheroids and $4\text{--}5\%$ for helicoidal particles.

\paragraph{Uncorrelated (random) closure error}

We first define single–particle error levels using the test dataset. Let
\begin{equation}
  \langle \cdot \rangle_{\mathrm{ds}}
\end{equation}
denote an average over dataset samples $j=1,\dots,N_{\mathrm{ds}}$. We introduce the RMS magnitude of the ideal single–particle stresslet as
\begin{equation}
  S_{\mathrm{typ}}^2 := \big\langle \|\mathbf{S}_\infty\|_F^2 \big\rangle_{\mathrm{ds}},
\end{equation}
and the RMS closure error level
\begin{equation}
  e_{\mathrm{cl}}^2 
  := \frac{\big\langle \|\delta\mathbf{S}\|_F^2 \big\rangle_{\mathrm{ds}}}
           {\big\langle \|\mathbf{S}_\infty\|_F^2 \big\rangle_{\mathrm{ds}}}.
\end{equation}
By construction this implies
\begin{equation}
  \big\langle \|\delta\mathbf{S}\|_F^2 \big\rangle_{\mathrm{ds}}^{1/2}
  = e_{\mathrm{cl}}\,S_{\mathrm{typ}}.
\end{equation}

In a given suspension with $N_p$ particles the ideal particle stress is
\begin{equation}
  \boldsymbol{\sigma}_p^\infty
  = \frac{1}{V}\sum_{n=1}^{N_p} \mathbf{S}_\infty^{(n)}.
\end{equation}
We assume that the single–particle stresslets $\mathbf{S}_\infty^{(n)}$ in the suspension are drawn from the same distribution as in the dataset and that, in the rheology–producing direction (simple shear, non–isotropic orientations), their contributions add coherently up to order–one factors. Under these assumptions the typical magnitude of the ideal particle stress is
\begin{equation}
  \|\boldsymbol{\sigma}_p^\infty\|
  \sim \frac{N_p}{V}\,S_{\mathrm{typ}}.
\end{equation}

The closure error in the particle stress is
\begin{equation}
  \Delta\boldsymbol{\sigma}_p
  = \boldsymbol{\sigma}_p^{\mathrm{cl}} - \boldsymbol{\sigma}_p^\infty
  = \frac{1}{V}\sum_{n=1}^{N_p} \delta\mathbf{S}^{(n)}.
\end{equation}
We now consider an ensemble of such suspensions and denote the corresponding expectation by $\langle\cdot\rangle_{\mathrm{ens}}$. In the random–error scenario we make two assumptions:
\begin{align}
  \big\langle \|\delta\mathbf{S}^{(n)}\|_F^2 \big\rangle_{\mathrm{ens}}
  &= \big\langle \|\delta\mathbf{S}\|_F^2 \big\rangle_{\mathrm{ds}}
  &&\text{for all } n, \label{eq:ass_same_stats} \\
  \big\langle \delta\mathbf{S}^{(n)} : \delta\mathbf{S}^{(m)} \big\rangle_{\mathrm{ens}}
  &\approx 0
  &&\text{for } n\neq m, \label{eq:ass_uncorr}
\end{align}
that is, each particle error has the same second moment as in the dataset and different particles have uncorrelated errors.

Using \eqref{eq:ass_same_stats}–\eqref{eq:ass_uncorr} one finds
\begin{equation}
  \big\langle \|\Delta\boldsymbol{\sigma}_p\|_F^2 \big\rangle_{\mathrm{ens}}
  \approx \frac{N_p}{V^2}\,
  \big\langle \|\delta\mathbf{S}\|_F^2 \big\rangle_{\mathrm{ds}}
  = \frac{N_p}{V^2}\,e_{\mathrm{cl}}^2 S_{\mathrm{typ}}^2,
\end{equation}
so that the RMS magnitude of the random error is
\begin{equation}
  \|\Delta\boldsymbol{\sigma}_p\|_{\mathrm{rand}}
  := \big\langle \|\Delta\boldsymbol{\sigma}_p\|_F^2 \big\rangle_{\mathrm{ens}}^{1/2}
  \approx \frac{\sqrt{N_p}}{V}\,e_{\mathrm{cl}}\,S_{\mathrm{typ}}.
\end{equation}
Combining this with the estimate for $\|\boldsymbol{\sigma}_p^\infty\|$ yields the scaling
\begin{equation}
  e_{\sigma}^{\mathrm{rand}}
  \equiv \frac{\|\Delta\boldsymbol{\sigma}_p\|_{\mathrm{rand}}}
              {\|\boldsymbol{\sigma}_p^\infty\|}
  \sim \frac{e_{\mathrm{cl}}}{\sqrt{N_p}},
\end{equation}
up to order–one prefactors. Thus, under the random–error assumptions, the relative rheology error contributed by the closure decays as $N_p^{-1/2}$ with the number of particles.

For typical quasi--dilute FCM simulations with $N_p$ in the range $10^2$–$10^3$, this gives
\begin{align}
  e_{\sigma}^{\text{rand,\,sph}} 
  &\sim \frac{1\%}{\sqrt{10^2\text{--}10^3}}
   \approx 0.03\text{--}0.1\%, \\
  e_{\sigma}^{\text{rand,\,hel}} 
  &\sim \frac{5\%}{\sqrt{10^2\text{--}10^3}}
   \approx 0.16\text{--}0.5\%.
\end{align}
The random component of the closure error is therefore strongly suppressed at the rheology level for realistic particle counts. For helicoidal particles, where the total closure error is dominated by a systematic BEM discretisation bias, these estimates should be viewed as upper bounds on the random contribution, which in practice is controlled by the smaller neural network component.

\paragraph{Summary}
By combining the micro–level analysis of the BEM–surrogate closure with simple error–propagation arguments, we obtain the following picture. For spheroids the total closure error on the deviatoric stresslet is of order $1\%$ (up to $2\%$ at the $95$th percentile), which directly translates into a $1\text{--}2\%$ uncertainty in the particle contribution to the bulk stress in the biased scenario and an even smaller error in the random scenario. For helicoidal particles the closure accuracy is limited by the BEM discretisation to about $4\text{--}5\%$ (up to $6\text{--}8\%$), with the neural network contributing only an $\mathcal{O}(1\%)$ correction. These error levels are well within what is typically considered quantitatively accurate for FCM-based rheology predictions and confirm that the surrogate does not introduce a new dominant source of uncertainty compared to the underlying microhydrodynamic solver.

\section{Conclusions and outlook}
\label{sec:conclusions}
We introduced a neural-network closure that emulates single-particle boundary-element (BEM) hydrodynamics and can be used as a drop-in replacement for analytical Fax\'en-type relations within the force-coupling method (FCM), thereby extending FCM-style many-particle simulations to genuinely three-dimensional complex shapes. The surrogate is trained offline on validated regularized-Stokeslet BEM data and predicts the deviatoric stresslet and rigid-body rotation (and, for helicoidal particles, the chiral thrust) from the local linearized flow and particle orientation. For the best models, the held-out test-set median relative errors are below $1\%$ for the deviatoric stresslet and $\Omega$ for both spheroids and helicoidal particles, with similarly small errors for the chiral thrust of helices.

\paragraph{Assumptions and limitations}
The present framework is designed for (i) Stokes flow at the particle scale (negligible inertia, linear response to the local velocity gradient), (ii) rigid, neutrally buoyant, non-Brownian particles, and (iii) quasi-dilute suspensions where a single-particle closure dominates the stress response and many-body effects enter primarily through FCM hydrodynamic interactions. In particular, the surrogate does not yet account for near-contact lubrication, mechanical contacts, or other short-range corrections that become essential as $\phi$ increases and interparticle gaps shrink. Moreover, the surrogate is shape-specific: generating a new closure requires an offline BEM dataset for that geometry, and predictive reliability is not guaranteed outside the flow/orientation ranges represented in the training set.

\paragraph{Future perspectives}
A direct next step is to embed the surrogate into an FCM solver and quantify end-to-end accuracy and speedups in canonical rheology setups (periodic shear and extensional flows), followed by coupling to general background flows obtained from CFD (spatially varying $\nabla \bar{\boldsymbol u}$) in complex geometries. Beyond this baseline, several extensions are natural: (i) augmenting FCM with lubrication and contact models while retaining the surrogate closure for the far-field stresslet response; (ii) broadening training to include a richer set of linear flows (randomized $(\boldsymbol E,\boldsymbol W)$) and enforcing additional symmetry/constraint penalties; (iii) incorporating uncertainty quantification to propagate surrogate error into macroscopic rheological observables; and (iv) expanding the closure to cover additional physics (finite-$Re$ corrections, Brownian rotation, or fluid--structure interaction for deformable particles) once appropriate training data and validation benchmarks are established.

\paragraph{Potential applications with FCM coupling}
When coupled to FCM, the surrogate enables statistically converged, parameter-sweep studies for suspensions of complex particles that are currently inaccessible with fully resolved simulations. High-impact directions include microfluidic chiral separation and sorting, fast prediction of viscosity and normal-stress trends in quasi-dilute suspensions containing non-spheroidal fillers, chiral colloids and functional self-assembly.

Overall, the results support the central premise of the work: once the expensive geometry-resolved hydrodynamics are learned offline, FCM-scale simulations retain their favorable scaling while gaining access to a broad class of three-dimensional particle shapes through a data-driven, symmetry-aware closure.

\section*{Acknowledgments}
M.L. was supported by Swedish Research Council Grant No. 2022-03032. The numerical computations were enabled by resources provided by the National Academic Infrastructure for Supercomputing in Sweden (NAISS) at the PDC Center for High Performance Computing, KTH Royal Institute of Technology, Sweden, partially funded by the Swedish Research Council through grant agreement no. 2022-06725.
The author is thankful to dr. Luca Manzari for the inspiring and helpful scientific discussions.

%%%%%%%%%%%%%%%%%%%%%%%%%%%%%%%%%%%%%%%%%%%%%%%%%%%%%%%%%%%%%%%%%%%%%%%%%%%%%%%%%%%%%%%%%%%%%%%%%
%\begin{thebibliography}{00}

%% For numbered reference style
%% \bibitem{label}
%% Text of bibliographic item

% \bibitem{lamport94}
%   Leslie Lamport,
%   \textit{\LaTeX: a document preparation system},
%   Addison Wesley, Massachusetts,
%   2nd edition,
%   1994.

% \bibliographystyle{unsrt}
% \bibliography{biblio}

% \end{thebibliography}
\end{document}